\documentclass[a4paper,12pt]{article}
\usepackage[paper=a4paper,left=25mm,right=25mm,top=25mm,bottom=25mm]{geometry}

\usepackage[english]{babel}
\usepackage[utf8]{inputenc}
\usepackage{longtable}
\usepackage{comment}
\usepackage{graphicx}
\usepackage{xcolor}
\usepackage{setspace}
\usepackage{soul}

\usepackage{pdflscape}

\usepackage{chngcntr}
\usepackage{lscape}
\usepackage{makecell}

\usepackage[skip=5pt, font=footnotesize, labelfont=footnotesize, textfont=footnotesize, tableposition=bottom]{caption}

\renewcommand{\footnotesize}{\small}

\usepackage{booktabs} 
\usepackage{colortbl} 
\usepackage{array} 

\usepackage{csquotes}

\usepackage{array}
\usepackage{arydshln} 
\usepackage{amsmath}

\usepackage[bottom]{footmisc}
\usepackage{scalerel}
\usepackage{tikz}

\usepackage{amssymb}

\usepackage{authblk}

\usepackage{orcidlink}
\usepackage{url}
\usepackage{xcolor} 

\usepackage[style=nature,maxnames=99,minnames=99]{biblatex}

\DeclareFieldFormat{urldate}{}
\renewbibmacro*{urldate}{}  

\DeclareFieldFormat{isbn}{}
\renewbibmacro*{isbn}{}  

\AtEveryBibitem{%
  \clearfield{month}%
  \clearfield{note}%
  \clearfield{issn}%
}

\definecolor{JournalBlue}{RGB}{0, 12, 146}

\usepackage{hyperref}

\hypersetup{
pdfborder={0 0 0},
    colorlinks=true,
    linkcolor=JournalBlue,  
    citecolor=JournalBlue,  
    urlcolor=JournalBlue    
}

\usepackage{tabularray}
\usepackage{float}
\usepackage{graphicx}
\usepackage{rotating}
\usepackage[normalem]{ulem}
\UseTblrLibrary{booktabs}
\NewTableCommand{\tinytableDefineColor}[3]{\definecolor{#1}{#2}{#3}}

\usepackage{authblk}

\title{\Large Gender and Discipline Shape Length, Content and Tone \\ of Grant Peer Review Reports}

\date{\normalsize Version: \today}

\author[1]{Stefan Müller\,\orcidlink{0000-0002-6315-4125}\thanks{\small Corresponding author:  
Stefan Müller (\href{mailto:stefan.mueller@ucd.ie}{stefan.mueller@ucd.ie})}}
\author[2]{Gabriel Okasa\,\orcidlink{0000-0002-3573-7227}} 
\author[2]{\\ Michaela Strinzel\,\orcidlink{0000-0003-3181-0623}} 
\author[2]{Anne Jorstad\,\orcidlink{0000-0002-6438-1979}}
\author[2]{\\  Katrin Milzow\,\orcidlink{0009-0002-8959-2534}}
\author[3,4]{Matthias Egger\,\orcidlink{0000-0001-7462-5132}}

\affil[1]{\small University College Dublin}
\affil[2]{\small Swiss National Science Foundation}
\affil[3]{\small University of Bern}
\affil[4]{\small University of Bristol}

\begin{document}

\maketitle

\vspace{-0.5cm}

\begin{abstract}
\noindent
Peer review by experts is central to the evaluation of grant proposals, but little is known about how gender and disciplinary differences shape the content and tone of grant peer review reports. We analyzed 39,280 review reports submitted to the Swiss National Science Foundation between 2016 and 2023, covering 11,385 proposals for project funding across 21 disciplines from the Social Sciences and Humanities (SSH), Life Sciences (LS), and Mathematics, Informatics, Natural Sciences, and Technology (MINT). Using supervised machine learning, we classified over 1.3 million sentences by evaluation criteria and sentiment. Reviews in SSH were significantly longer and more critical, with less focus on the applicant’s track record, while those in MINT were more concise and positive, with a higher focus on the track record, as compared to those in LS. Compared to male reviewers, female reviewers write longer reviews that more closely align with the evaluation criteria and express more positive sentiments. Female applicants tend to receive reviews with slightly more positive sentiment than male applicants. Gender and disciplinary culture influence how grant proposals are reviewed--shaping the tone, length, and focus of peer review reports. These differences have important implications for fairness and consistency in research funding.
\end{abstract}

\medskip


\pagebreak

\doublespacing

\setcounter{page}{1}


\section*{Introduction}

Peer review is the cornerstone of the scientific review process, aiming to ensure fair allocation of funding, based on merit, scientific quality and potential. Typically, experienced external reviewers provide structured assessments of proposals that inform the recommendations made by evaluation panels. The process of grant peer review has, however, been criticized for many years \cite{bornmann2011}. Reviewers often remain anonymous, applicants may not see full review reports, and the weight given to the external reports in final funding decisions is generally unclear \cite{jerrim2023, langfeldt2025,nicholls1999}. Compared to journal article review, researchers tend to have less trust in the grant review process \cite{langfeldt2024}. Moreover, peer review of both articles and grant proposals may be influenced by the gender and other characteristics of applicants and reviewers, potentially introducing bias \cite{lee2013,tamblyn2018, huber2022}. Approaches to peer review may vary across research areas or disciplines. Disciplinary differences and cultures shape how proposals are reviewed, with different fields emphasizing the applicants’ track record, methodological rigor, or originality. Finally, the well-documented low inter-rater reliability in peer review underscores the difficulty of achieving consistent evaluations across experts \cite{jayasinghe2001, cicchetti1991}.

Although billions of research dollars are distributed competitively using peer review, the process remains under-researched. While journal peer review and editorial practices have been the focus of a growing body of empirical research \cite{ghosal2022,verharen2023,severin2023,luo2021,erosheva2020}, the evidence on grant peer review is scarce. Simon Wessely has argued that studying grant peer review may be more important than studying publication practices: whereas published articles reflect completed research, rejected grant proposals often represent studies that will never be done \cite{wessely1998}. Funding success also has significant impacts on academic careers. An analysis based on funding outcomes at the Netherlands Organization of Scientific Research shows that applicants just above the funding threshold accumulated more than twice as much funding in subsequent years as applicants just below the threshold \cite{bol2018}. Decisions based on grant peer review can have lasting effects on research and careers. Studying the process is essential to ensure it is as fair and as transparent as possible.

The emergence of Large Language Models (LLMs) \cite{tunstall2022} has allowed increasingly sophisticated analyses of peer review reports \cite{kuznetsov2024}. Beyond thematic analysis, LLMs can assess the tone or sentiment of reviews by quantifying whether the language used is supportive, neutral, or critical \cite{liu2020}. Sentiment analysis can help funding agencies detect implicit positivity or negativity that might not be evident from review scores. We defined key content categories and trained four annotators to assign categories to 3,000 randomly sampled sentences from peer review reports submitted to the Swiss National Science Foundation’s (SNSF) project funding scheme \cite{okasa2024}. Using these annotations, we fine-tuned six transformer machine learning models. After demonstrating that classifiers performed reliably \cite{okasa2024}, we applied the models to over 1.3 million sentences from almost 40,000 peer review reports. This study examines how the length, content, and sentiment of peer review reports vary across fields and disciplines, as well as by the gender of applicants and reviewers. It is among the first to analyze the structure and textual content of grant peer review reports, extending research beyond numerical scores and ratings. In doing so, it offers novel insights for reviewers, applicants, and funding agencies alike.

\section*{Results}

We present the results in several steps. We start with descriptive statistics on the characteristics of reviews, reviewers and applicants. Next, we focus on the differences in the length of reviews across disciplines, the gender of reviewers, and the gender of the corresponding applicants. Afterwards, using descriptive and mixed-effects regression analyses, we assess how content related to the SNSF evaluation criteria, and the sentiment of peer review reports differ across disciplines and gender. Gender is self-reported, and during the period of analysis, the SNSF recorded only two categories for gender: female or male. Moreover, we report on the role of the review score, and on robustness analyses. 

\begin{table}[!h]
\centering \footnotesize
\resizebox{0.99\textwidth}{!}{
\begin{tabular}{lcccc}
\toprule 
& \textbf{LS} & \textbf{MINT} & \textbf{SSH} & \textbf{Total} \\
\midrule
\textbf{Reviewers} & \textbf{15,006 (100\%)} & \textbf{15,441 (100\%)} & \textbf{8,833 (100\%)} & \textbf{39,280 (100\%)} \\
\addlinespace
\multicolumn{5}{l}{\textbf{Gender of Reviewer}} \\
\quad Male & 11,525 (76.8\%) & 13,555 (87.8\%) & 5,433 (61.5\%) & 30,513 (77.7\%) \\
\quad Female & 3,481 (23.2\%) & 1,886 (12.2\%) & 3,400 (38.5\%) & 8,767 (22.3\%) \\
\addlinespace
\multicolumn{5}{l}{\textbf{Country of Affiliation}} \\
\quad United States of America & 3,783 (25.2\%) & 3,371 (21.8\%) & 2,051 (23.2\%) & 9,205 (23.4\%) \\
\quad Germany & 1,187 (7.9\%) & 1,443 (9.3\%) & 805 (9.1\%) & 3,435 (8.7\%) \\
\quad Great Britain and N. Ireland & 1,072 (7.1\%) & 1,178 (7.6\%) & 1,072 (12.1\%) & 3,322 (8.5\%) \\
\quad France & 723 (4.8\%) & 866 (5.6\%) & 247 (2.8\%) & 1,836 (4.7\%) \\
\quad Italy & 529 (3.5\%) & 929 (6.0\%) & 282 (3.2\%) & 1,740 (4.4\%) \\
\quad Canada & 513 (3.4\%) & 583 (3.8\%) & 341 (3.9\%) & 1,437 (3.7\%) \\
\quad Netherlands & 583 (3.9\%) & 364 (2.4\%) & 419 (4.7\%) & 1,366 (3.5\%) \\
\quad Australia & 527 (3.5\%) & 387 (2.5\%) & 278 (3.1\%) & 1,192 (3.0\%) \\
\quad Other & 2,730 (18.2\%) & 3,722 (24.1\%) & 1,344 (15.2\%) & 7,796 (19.8\%) \\
\quad Unknown & 3,359 (22.4\%) & 2,598 (16.8\%) & 1,994 (22.6\%) & 7,951 (20.2\%) \\
\addlinespace
\midrule
\textbf{Applicants} & \textbf{4,264 (100\%)} & \textbf{4,077 (100\%)} & \textbf{3,044 (100\%)} & \textbf{11,385 (100\%)} \\
\addlinespace
\multicolumn{5}{l}{\textbf{Gender}} \\
\quad Male & 3,074 (72.1\%) & 3,401 (83.4\%) & 1,916 (62.9\%) & 8,391 (73.7\%) \\
\quad Female & 1,190 (27.9\%) & 676 (16.6\%) & 1,128 (37.1\%) & 2,994 (26.3\%) \\
\addlinespace
\multicolumn{5}{l}{\textbf{Age}} \\
\quad Mean (SD) & 48.2 (7.5) & 46.9 (8.2) & 48.2 (7.9) & 47.7 (7.9) \\
\addlinespace
\multicolumn{5}{l}{\textbf{Affiliation}} \\
\quad Cantonal University & 3,383 (79.3\%) & 1,507 (37.0\%) & 2,010 (66.0\%) & 6,900 (60.6\%) \\
\quad ETH Domain & 544 (12.8\%) & 2,150 (52.7\%) & 183 (6.0\%) & 2,877 (25.3\%) \\
\quad UAS/UTE & 66 (1.5\%) & 221 (5.4\%) & 547 (18.0\%) & 834 (7.3\%) \\
\quad Other & 167 (3.9\%) & 160 (3.9\%) & 279 (9.2\%) & 606 (5.3\%) \\
\quad Hospital & 86 (2.0\%) & 2 (0.0\%) & 3 (0.1\%) & 91 (0.8\%) \\
\quad Unknown & 18 (0.4\%) & 37 (0.9\%) & 22 (0.7\%) & 77 (0.7\%) \\
\addlinespace
\multicolumn{5}{l}{\textbf{Academic Rank}} \\
\quad Full Professor & 1,098 (25.8\%) & 1,173 (28.8\%) & 1,254 (41.2\%) & 3,525 (31.0\%) \\
\quad Associate Professor & 858 (20.1\%) & 534 (13.1\%) & 468 (15.4\%) & 1,860 (16.3\%) \\
\quad Assistant Professor & 416 (9.8\%) & 446 (10.9\%) & 312 (10.2\%) & 1,174 (10.3\%) \\
\quad UAS/UTE Professor & 27 (0.6\%) & 84 (2.1\%) & 328 (10.8\%) & 439 (3.9\%) \\
\quad Senior Postdoc/Group Leader & 1,550 (36.4\%) & 1,540 (37.8\%) & 599 (19.7\%) & 3,689 (32.4\%) \\
\quad Other Professorship & 290 (6.8\%) & 274 (6.7\%) & 76 (2.5\%) & 640 (5.6\%) \\
\bottomrule
\end{tabular}
}
\caption{\textit{Distribution of review reports across research domains and reviewer and applicant characteristics. }\textit{Notes}: UAS/UTE: University of Applied Sciences or University of Teacher Education; Other professorships include honorary, titular and visiting professors; LS: Life Sciences; MINT: Mathematics, Informatics, Natural Sciences, Technology; SSH: Social Sciences and Humanities.}
\label{tab_01}
\end{table}

\subsection*{Characteristics of peer reviewers and applicants}

Our analysis covers peer review reports for the 14 project funding calls between October 2016 and April 2023. Out of a total of 44,012 reports written in English, German, French or Italian, we kept 39,280 English language reports (89.2\% of sample), corresponding to 11,385 proposals with about 3.5 reviews per proposal. These reports comprised 1,304,621 sentences and 30,477,479 words. Table 1 presents the distribution of review reports by research domain and reviewer and applicant characteristics. The Mathematics, Informatics, Natural Sciences, Technology (MINT) and Life Sciences (LS) domains account for around 15,000 reviews each, while the Social Sciences and Humanities (SSH) had fewer (just under 9,000), in line with the larger number of proposals submitted in MINT and LS. Across all fields of science, there were fewer female reviewers than male reviewers. This was particularly pronounced in MINT (12.2\%), followed by LS (23.2\%), with the highest contribution of female reviewers in SSH (38.5\%). About a quarter of these reviews was submitted by scholars based in the United States of America (23.4\%), followed by Germany (8.7\%), Great Britain and Northern Ireland (8.5\%), and France (4.7\%). In line with SNSF policy, only 88 reviews were submitted by experts from a Swiss institution. For 7,951 (20.2\%) reviews, information on the reviewer’s country was not available.

We observe similar gender differences for applicants. Few women applied for funding in MINT (16.6\% female applicants), while LS and SSH had more female applicants (27.9\% and 37.1\% female, respectively). The mean age of applicants was 47.7 years, and most (60.6\%) worked at a Swiss cantonal university, followed by applicants based in the ETH domain (25.3\%), i.e., at one of the two Federal Institutes of Technology (ETH Zurich or EPFL Lausanne) or at one of the four ETH research institutions. In MINT, but not in LS or SSH, applications from ETH dominated (Table \ref{tab_01}). Almost one in three applicants was a full professor (31.0\%), followed by associate professors (16.3\%) and assistant professors (10.3\%). About a third of applicants were senior postdocs or group leaders (32.4\%).

\subsection*{Review length}

\begin{figure}[ht]
\centering
\includegraphics[width=0.95\textwidth]{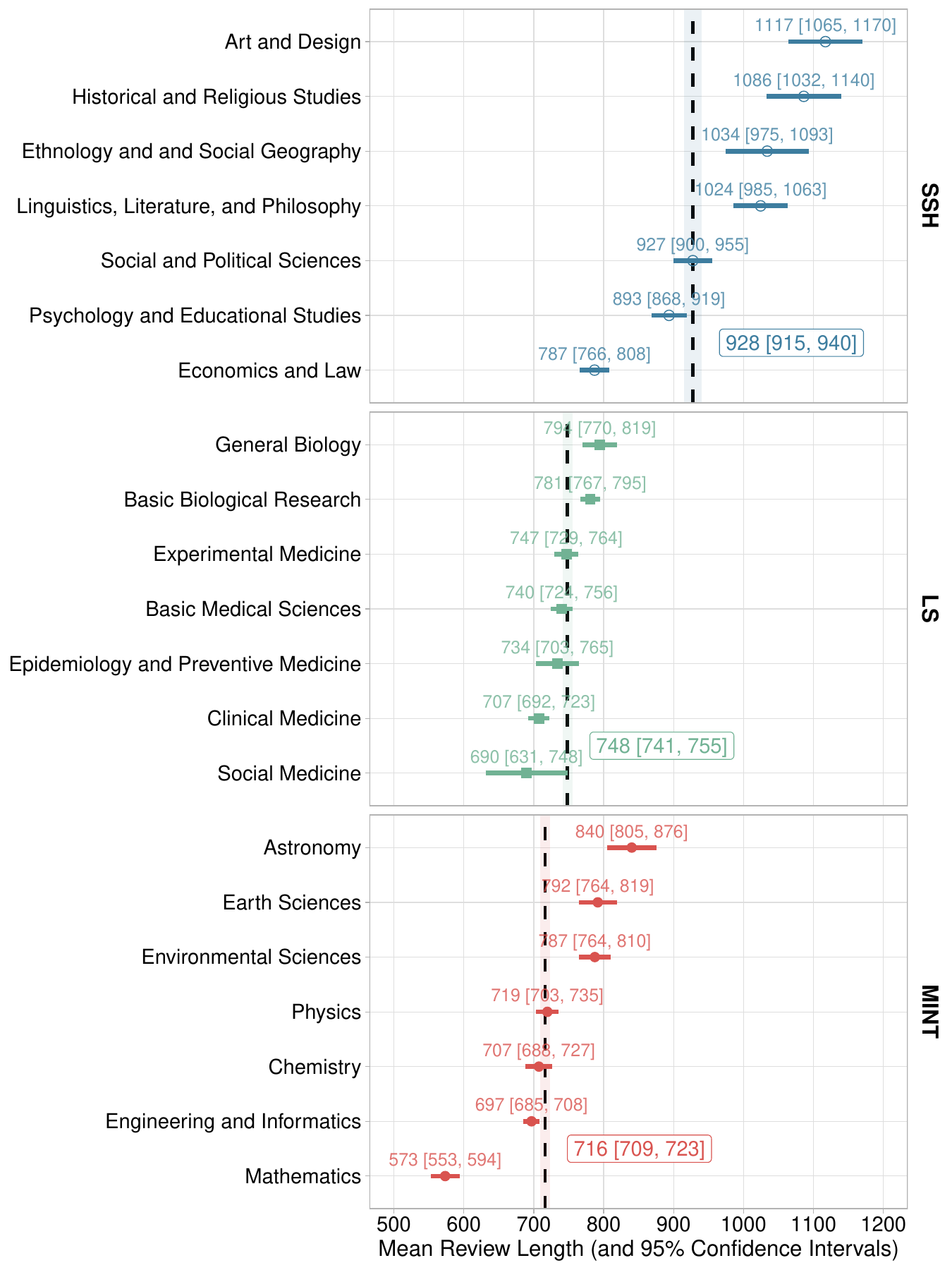}
\caption{\textit{Average review length conditionally on the research domain}. Dots show the mean length (in words), horizontal bars the 95\% confidence intervals. SSH: Social Sciences and Humanities; LS: Life Sciences; MINT: Mathematics, Informatics, Natural Sciences, Technology. The category Art and Design includes musicology, theatre and film studies and architecture. Social and Political Sciences include medical sociology, social work and media and communication studies. Astronomy includes astronomy, astrophysics, and space sciences. The names of some of the disciplines have been shortened to fit in the figure; the full names can be found in Table \ref{tab_s02}.} \label{fig_01}
\end{figure}

The word count of reviews differed across the 21 SNSF disciplines (Fig.\,\ref{fig_01}). With a mean of 928 words, reviews in the SSH are considerably longer than reviews in the LS and MINT. We observe further differences within each research domain. With 1,117 words, reviews in the discipline \textit{Art and Design} receive the longest reviews, closely followed by \textit{Historical and Religious Studies} (1,086 words). In contrast, the reviews submitted in \textit{Economics and Law} are shorter than 800 words. The mean length in LS varies from 690 words in \textit{Social Medicine} to 794 words in \textit{General Biology}. By far the shortest reviews are submitted for proposals in \textit{Mathematics}: the mean review length was only 573 words. 

\pagebreak
\clearpage

\begin{figure}[ht]
\centering
\includegraphics[width=1\textwidth]{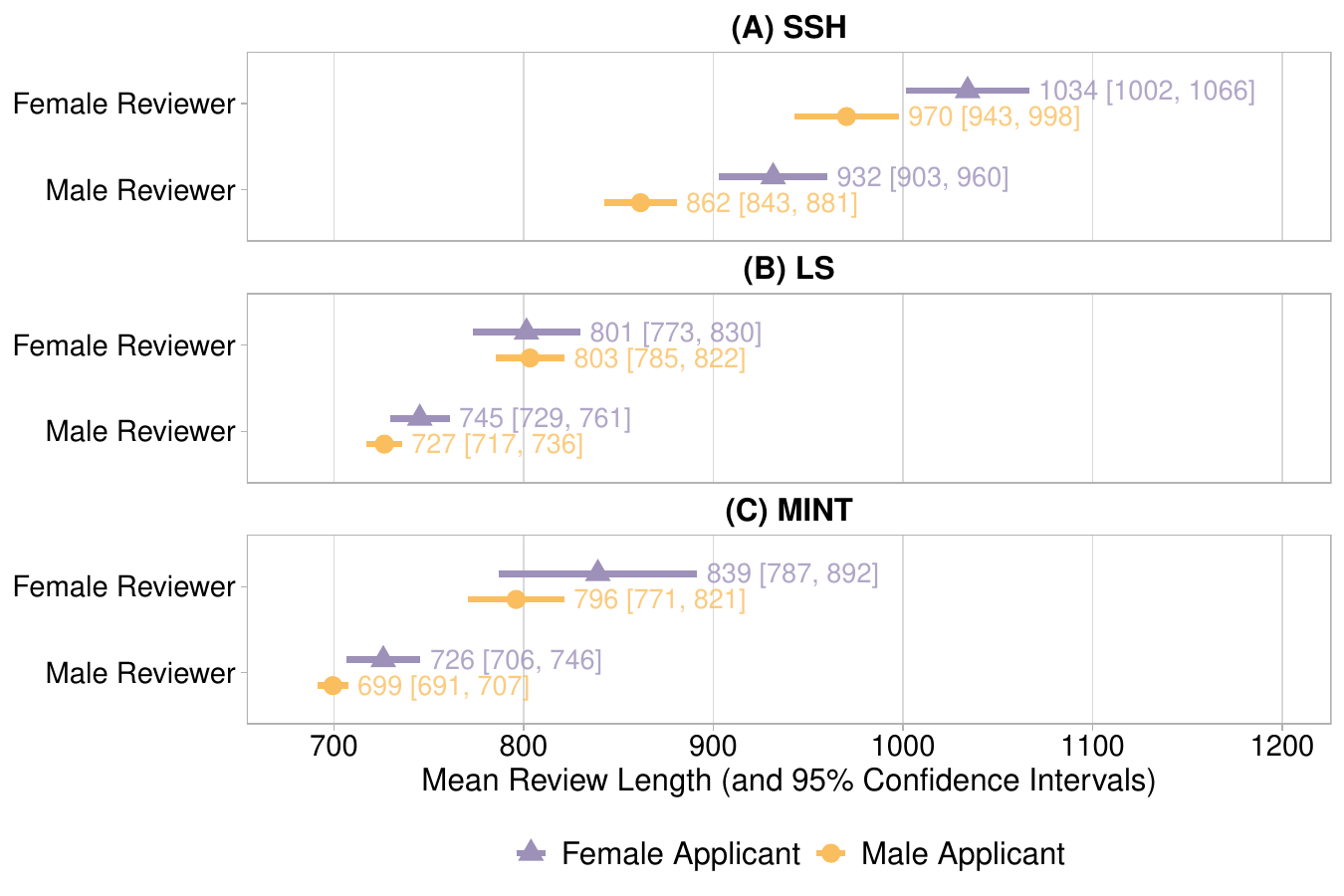}
\caption{\textit{Average review length by the gender of reviewers and gender of applicants}. Dots show mean length (in words), horizontal bars depict 95\% confidence intervals. SSH: Social Sciences and Humanities; LS: Life Sciences; MINT: Mathematics, Informatics, Natural Sciences, Technology.} \label{fig_02}
\end{figure}

Female reviewers submitted longer reviews than male reviewers in all three research domains (Fig.\,\ref{fig_02}). The difference was largest in the SSH. In SSH and MINT, but not in the LS, female applicants received, on average, longer reviews from both female and male reviewers. This difference was again most pronounced for the SSH. Of note, the gender of reviewers is more strongly associated with review length than the gender of applicants (Supplementary Materials, Fig.\,\ref{fig_s01}).

Analyzing review length by the overall grade given to the proposal by the peer reviewers, we found an inverted U-shaped relationship between the grade and the review’s word count (Supplementary Materials, Fig.\,\ref{fig_s02}). Peer review reports for proposals graded at the bottom of the scale as low quality received shorter reviews than higher graded proposals, with the longest reports submitted for proposals with intermediate grades.

\subsection*{Review content}

To investigate content, we focus on the four SNSF evaluation criteria—(i) Track Record; (ii) Relevance, Originality, Topicality; (iii) Suitability of Methods, and (iv) Feasibility of Project—as well as the sentiment of peer review reports (Positive and Negative).

Table \ref{tab:02} presents the 25 most predictive terms for each category from keyness analyses \cite{severin2023,zollinger2024}. In the evaluation criteria, sentences labeled as \textit{Track Record} frequently included terms such as “unk” (a placeholder for anonymized “unknown” names of persons), “expertise,” “applicant,” “publications,” “track\_record,” “professor,” “cv,” and “excellent\_track\_record,” highlighting references to professional background and academic productivity. The category of \textit{Relevance, Originality, Topicality} was marked by words like “original,” “innovative,” “highly\_relevant,” “scientific\_relevance,” “broader\_impact,” and “topical,” pointing to novelty, potential, and scientific contribution. Sentences associated with \textit{Suitability of Methods} often contained terms such as “proposed\_methods,” “chosen\_methods,” “methodology,” “well\_suited,” “methods\_described,” and “sample\_size,” reflecting an emphasis on the appropriateness and clarity of research methods. For \textit{Feasibility of Project}, frequent terms included “feasibility,” “milestones,” “timeline,” \phantom{...} “planned\_duration,” “available\_resources,” and “highly\_feasible,” indicating attention to timelines, planning, and resource allocation.

In terms of sentiment, \textit{Positive} was reflected in terms such as “excellent,” “original,” “outstanding,” “applicant,” “proposed\_project,” “highly\_relevant,” and “innovative,” which convey approval, strength, and merit. In contrast, \textit{Negative} was associated with terms like “lack,” “unclear,” “difficult,” “missing,” “weakness,” “concern,” and “vague,” pointing to criticisms, uncertainties, or limitations. These distinctions in language use provide evidence for the validity of the content categories, identified through human annotations and machine learning.

\pagebreak
\clearpage

\begin{table}[!h]
\centering
\begingroup\small
\begin{tabular}{p{0.35\textwidth}p{0.6\textwidth}}
  \toprule
\textbf{Category} & \textbf{Predictive Terms} \\ 
  \hline
  \multicolumn{2}{c}{\textit{\textbf{Evaluation Criteria}}}\\
  \hdashline 
Track Record & unk, expertise, applicant, dr, publications, experience, track\_record, field, published, prof, pi, carry, team, journals, papers, expert, professor, university, phd, excellent\_track\_record, research, publication\_record, career, excellent, cv \\ 
   \hdashline 
Relevance, Originality, Topicality & original, originality, novel, timely, topical, topicality, scientific\_relevance, topic, relevant, relevance, innovative, highly\_relevant, understanding, novelty, interesting, potential, discipline, important, field, scientifically\_relevant, exciting, project, broader\_impact, proposed\_project, interest \\ 
   \hdashline 
Suitability of Methods & methods, suitable, appropriate, suitability, proposed\_methods, methodology, methods\_proposed, well\_suited, suited, method, chosen\_methods, answering, adequate, questions\_set, methods\_chosen, feasible, chosen, choice, approach, proposed\_methodology, sample\_size, answer, methods\_described, used, use \\ 
   \hdashline 
Feasibility of Project & feasible, feasibility, milestones, timeline, project, workload, budget, schedule, personnel, planned\_duration, proportionate, given\_time, reasonable, achievable, time, time\_frame, realistic, highly\_feasible, available\_resources, planned, ambitious, timeframe, milestones\_set, reached, timing \\ 
   \hdashline
   \multicolumn{2}{c}{\textit{\textbf{Sentiment}}}\\
   \hdashline 
Positive & field, project, excellent, unk, expertise, original, applicant, good, overall, proposed\_project, relevant, timely, outstanding, carry, highly\_relevant, research, team, feasible, strong, topical, interesting, track\_record, proposed\_research, innovative, topic \\ 
   \hdashline 
Negative & lack, however, unclear, clear, difficult, little, missing, weakness, concern, limited, bit, somewhat, seems, unfortunately, seem, vague, lacks, weaknesses, lacking, proposal, details, description, concerns, although, whether \\ 
   \bottomrule
\end{tabular}
\endgroup
\caption{\textit{Frequent predictive terms for each content category.} The results are based on keyness analyses using $\chi^2$ tests for each word or multi-word expression, comparing frequencies in sentences where a content characteristic was present (target group) with those where it was absent (reference group). The term ``unk'' is a placeholder for anonymized names of persons involved in a proposal.\label{tab:02}} 
\end{table}

\begin{figure}[ht]
\centering
\includegraphics[width=1\textwidth]{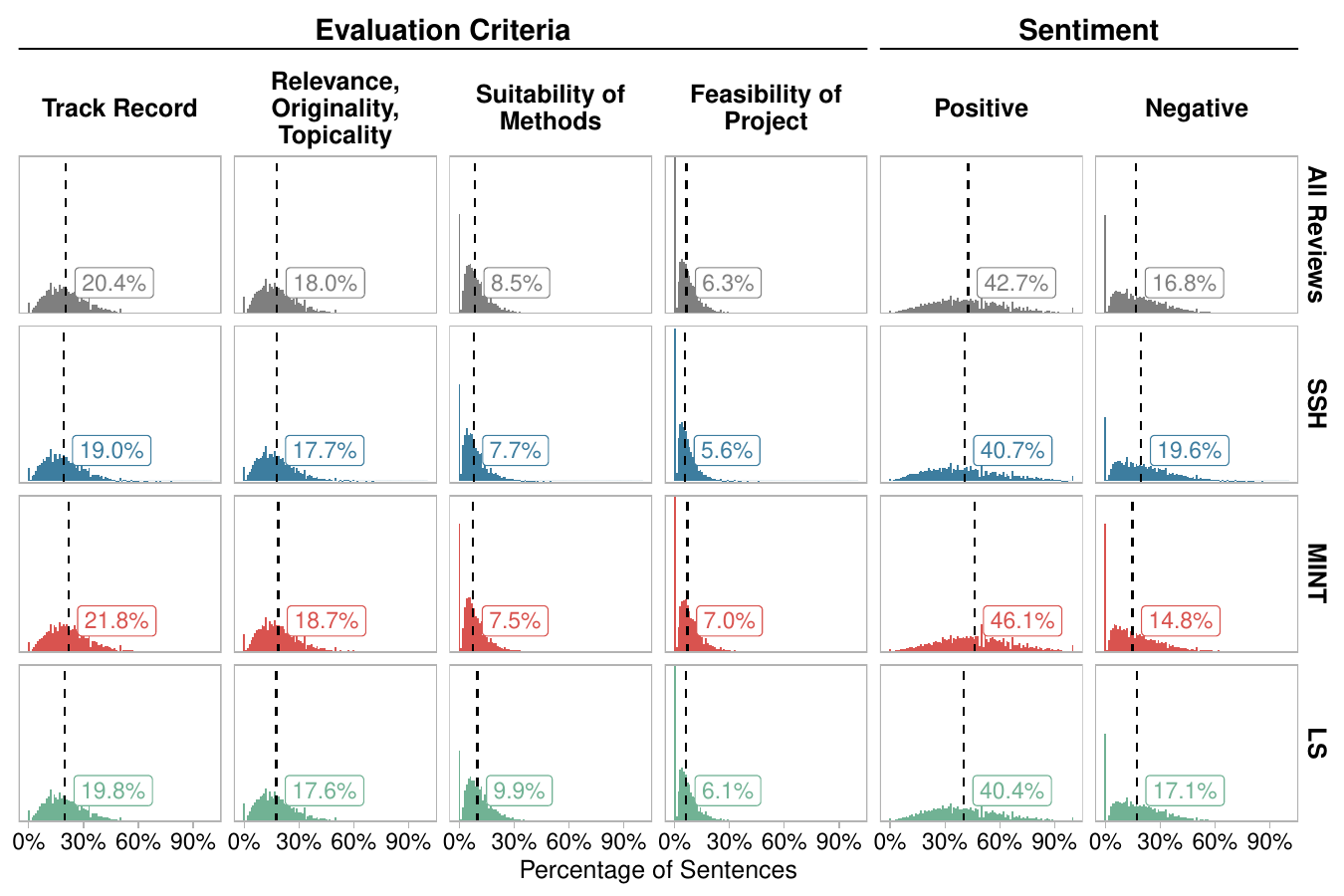}
\caption{\textit{Distribution of sentences in peer review reports allocated to four evaluation criteria and two sentiment categories for the full sample and by research domain}. A sentence could be allocated to no, one, or several categories. Vertical dashed lines and numbers in graphs depict the mean percentage of sentences after aggregating them to the level of reviews. SSH,  Social Sciences and Humanities; MINT, Mathematics, Informatics, Natural Sciences and Technology; LS, Life Sciences. } \label{fig_03}
\end{figure}

Fig.\,\ref{fig_03} presents the distribution of each content category across all reviews and by research domain. Several patterns emerge. First, the applicant’s \textit{Track Record} and the \textit{Relevance, Originality, and Topicality} of the proposal are the most frequently addressed criteria. Across all reviews, the track record accounts for an average of 20.4\% of sentences, while 18.0\% discuss the relevance, originality, and topicality of the proposed research.

A comparison across fields shows that reviews in MINT place greater emphasis on both the applicant’s track record and the scientific relevance of the proposal. \textit{Suitability of Methods} is discussed in 8.5\% of all sentences, with LS placing comparatively more emphasis on this criterion (9.9\% of sentences). \textit{Feasibility} is addressed in 6.3\% of sentences overall, with the SSH allocating slightly fewer sentences (5.6\%) to this criterion than LS (6.1\%) and MINT (7.0\%). Fig.\,\ref{fig_03} also shows that peer review reports are generally more positive than negative in tone. On average, 42.7\% of review sentences express a positive sentiment, while only 16.8\% are negative. Notably, reviews in the MINT fields are more positive on average than those in the other two domains.

\pagebreak

\begin{figure}[h!]
\centering
\includegraphics[width=0.99\textwidth]{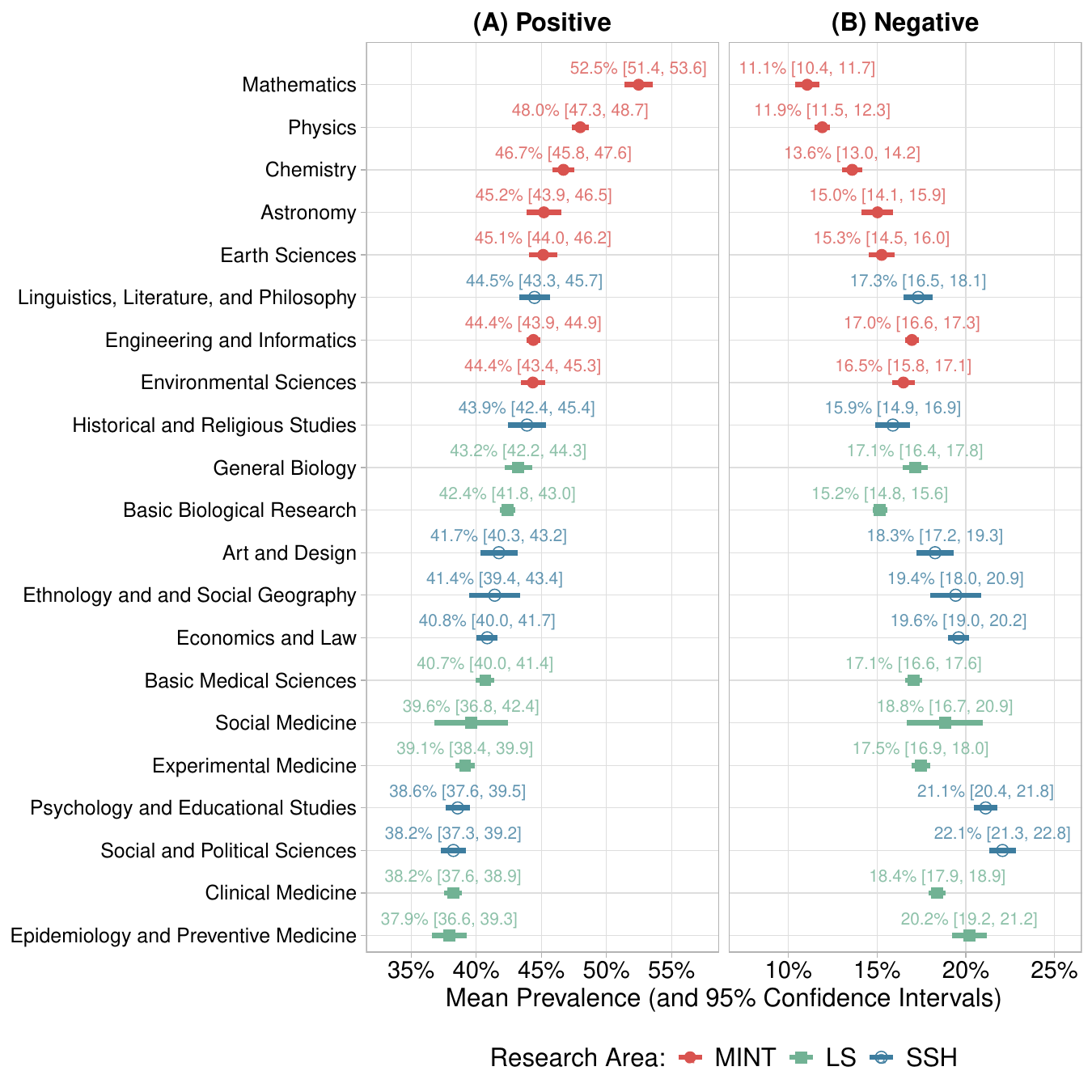}
\caption{\textit{Average prevalence of positive (A) and negative (B) statements (\% of review sentences) by discipline}. Horizontal bars show 95\% confidence intervals. Separate horizontal axis limits to allow for comparisons within each content category. Overall, reviews tend to be much more positive than negative. SSH: Social Sciences and Humanities; LS: Life Sciences; MINT: Mathematics, Informatics, Natural Sciences, Technology. The category Art and Design includes musicology, theatre and film studies and architecture. Social and Political Sciences include medical sociology, social work and media and communication studies. Astronomy includes astronomy, astrophysics, and space sciences. The names of some of the disciplines have been shortened to fit in with the figure; the full names can be found in Table \ref{tab_s02}. Corresponding figures showing the distribution of each category across all disciplines are provided in Supplementary Materials, Figs.\,\ref{fig_s03}--\ref{fig_s06}.} \label{fig_04}
\end{figure}

\pagebreak

Further stratifying the analysis by academic discipline shows that \textit{Mathematics} (25.6\%), \textit{Physics} (23.5\%), \textit{Astronomy} (22.5\%), and \textit{Chemistry} (22.4\%) place the most emphasis on the applicant’s track record (Supplementary Materials, Fig.\,\ref{fig_s03}). In contrast, disciplines within the social sciences and health fields, such as \textit{Social and Political Sciences} show the lowest prevalence (17.6\%). For relevance, originality, and topicality (Fig.\,\ref{fig_s04}), the differences across disciplines are smaller, with no strong clustering by field, suggesting a relatively consistent application of this criterion across disciplines. In contrast, the suitability of the methods shows larger differences (Fig.\,\ref{fig_s05}), with the LS, particularly \textit{Epidemiology and Preventive Medicine} and \textit{Social Medicine} emphasizing this criterion (11.8\% and 11.4\% of review sentences, respectively), compared to under 5.5\% in humanities-oriented SSH disciplines, such as \textit{Historical and Religious Studies}. Finally, the feasibility of the proposed research (Fig.\,\ref{fig_s06}) is discussed slightly more in MINT and LS disciplines than in SSH.

There were substantial differences in the tone of peer review reports across the 21 disciplines.  Reviews in all MINT disciplines are more positive on average than those in SSH and LS (Fig.\,\ref{fig_04}, Panel A). For example, in \textit{Mathematics}, more than half of the sentences in an average review (52.5\%) were positive, compared to only 37.9\% in \textit{Epidemiology and Preventive Medicine}. We observe similar patterns for the prevalence of negative sentences. Reviews in SSH disciplines, in particular \textit{Social and Political Sciences} as well as \textit{Psychology and Educational Studies}, were more negative (22.1\% and 21.1\%, respectively) than in the MINT disciplines. For example, only 11.1\% of sentences in reviews of \textit{Mathematics} proposals and 11.9\% of \textit{Physics} proposals were negative.

\subsection*{Regression models}

The results from multivariate mixed-effects regression models show that proposals reviewed by female experts include more content directly related to the evaluation criteria, even when adjusting for the length of reviews (Table \ref{tab:03}). Coefficients indicate percentage point differences relative to the reference category (male).

The evaluation criteria were covered more comprehensively by female reviewers than by male reviewers. In particular, the track record of the applicants, the relevance and originality of the project and its feasibility were covered in greater detail.  Further, female reviewers made fewer negative comments than male reviewers (difference -0.98, 95\% confidence intervals (CI) [--1.32,  --0.65]). Focusing on the gender of the applicants, female applicants received slightly more methodological comments than male applicants (difference 0.43, 95\% CI [0.25, 0.61]), with little evidence of differences for the other evaluation criteria. Further, female applicants received both slightly more positive and fewer negative remarks on average (difference 0.50, 95\% CI [0.02, 0.97] and -0.46, 95\% CI [-0.82, -0.10] percentage points, respectively).

\begin{table}
\centering 
\resizebox{0.99\textwidth}{!}{
\begin{talltblr}[         
caption={\textit{Predicting peer review content.} Coefficients from mixed-effects models show percentage point differences in prevalence relative to reference category. Models include control variables listed in Table \ref{tab_s01} and random intercepts for proposal IDs. 95\% confidence intervals in parentheses. Analysis based on 39,280 peer review reports on 11,385 proposals. Track Record: Applicant's Track Record; Rel., Orig., Topic: Relevance, Originality, Topicality; Methods: Suitability of Methods; Feasibility: Feasibility of Project.\label{tab:03}},
]                     
{          
colspec={Q[]Q[]Q[]Q[]Q[]Q[]Q[]},
cell{1}{2}={c=4,}{halign=c,},
cell{1}{6}={c=2,}{halign=c,},
column{1}={halign=l,},
column{2}={halign=c,},
column{3}={halign=c,},
column{4}={halign=c,},
column{5}={halign=c,},
column{6}={halign=c,},
column{7}={halign=c,},
row{2}={font=\fontsize{1em}{1.3em}\selectfont,},
row{1}={font=\fontsize{1em}{1.3em}\selectfont,},
row{3}={font=\fontsize{1em}{1.3em}\selectfont,},
row{4}={font=\fontsize{1em}{1.3em}\selectfont,},
row{5}={font=\fontsize{1em}{1.3em}\selectfont,},
row{6}={font=\fontsize{1em}{1.3em}\selectfont,},
row{7}={font=\fontsize{1em}{1.3em}\selectfont,},
row{8}={font=\fontsize{1em}{1.3em}\selectfont,},
row{9}={font=\fontsize{1em}{1.3em}\selectfont,},
row{10}={font=\fontsize{1em}{1.3em}\selectfont,},
}                     
\toprule
& Evaluation Criteria &  &  &  & Sentiment &  \\ \cmidrule[lr]{2-5}\cmidrule[lr]{6-7}
& \makecell{Track \\ Record} & \makecell{Rel., \\ Orig., \\ Topic.} & Methods & Feasibility & Positive & Negative \\ \midrule 
Reviewer: Female   & 1.00           & 0.76          & 0.48           & 0.60           & 0.59         & -0.98          \\
(ref.: Male) & [0.74, 1.26]   & [0.52, 1.01]  & [0.31, 0.64]   & [0.46, 0.74]   & [0.15, 1.04] & [-1.31, -0.65] \\[1.5ex]
Applicant: Female & 0.22           & -0.05         & 0.43           & 0.03           & 0.50         & -0.46          \\
(ref.: Male) & [-0.05, 0.48]  & [-0.31, 0.20] & [0.25, 0.61]   & [-0.12, 0.17]  & [0.02, 0.97] & [-0.82, -0.10] \\[1.5ex]
Domain: MINT        & 1.82           & 0.61          & -2.19          & 0.74           & 4.65         & -1.72          \\
(ref.: LS) & [1.52, 2.11]   & [0.33, 0.88]  & [-2.39, -1.99] & [0.58, 0.90]   & [4.13, 5.17] & [-2.12, -1.33] \\[1.5ex]
Domain: SSH          & -0.77          & 0.81          & -2.09          & -0.44          & 2.71         & 1.59           \\
(ref.: LS) & [-1.09, -0.45] & [0.50, 1.11]  & [-2.31, -1.87] & [-0.61, -0.26] & [2.14, 3.28] & [1.16, 2.02]   \\
\bottomrule
\end{talltblr}
}
\end{table}

There were also notable differences across the three research domains. Compared to LS, proposals in the MINT and SSH disciplines gave less emphasis to the suitability of methods (difference around --2 percentage points), and more emphasis to the relevance and originality of the research (differences 0.61, 95\% CI [0.33, 0.88] and 0.81, 95\% CI [0.50, 1.11] percentage points, respectively). Reviews of MINT projects emphasized the track record of applicants and the feasibility of the research more than LS reviewers (differences 1.82, 95\% CI [1.52, 2.11] and 0.74, 95\% CI [0.58, 0.90] respectively) whereas the opposite was the case for SSH (differences --0.77, 95\% CI [--1.09, --0.45] and --0.44, 95\% CI [--0.61, --0.26], respectively). Interestingly, MINT proposals received substantially more positive comments on average than LS projects (difference 4.65, 95\% CI [4.13, 5.17]) and fewer negative comments (--1.72, 95\% CI [--2.12, --1.33]), whereas SSH proposals received both more positive and more negative comments (differences 2.71, 95\% CI [2.14, 3.28] and 1.59, 95\% CI [1.16, 2.02], respectively).

\subsection*{Further analyses}

We conducted additional analyses to assess the robustness of the results presented above. When including an interaction term in the model (female reviewer $\times$ female applicant), we find that when both the reviewer and applicant are female, reviews include more positive and fewer negative comments. In contrast, there was little evidence for interactions between the genders of reviewers and applicants for the evaluation criteria (Supplementary Materials, Table \ref{tab:s03}). Stratifying this analysis by research area, it becomes clear that these results are driven by the LS (Supplementary Materials, Fig.\,\ref{fig_s07}).

When including the overall grades into the model (Table \ref{tab:s04} in the Supplementary Materials), higher grades were associated with an increase in text on all evaluation criteria except for the methods, which received less attention (difference per grade point increase --0.45, 95\% CI [--0.51, --0.39]). As could be expected, positive comments increased substantially with higher grades (9.60, 95\% CI [9.46, 9.73] percentage points per grade point increase) whereas negative comments decreased (--8.00, 95\% CI [--8.10, --7.91]). While grades strongly predict some evaluation criteria and sentiment, the observed differences across research areas and by the gender of reviewers and applicants remain.

To assess the robustness of our results with respect to the modelling assumptions, we relaxed the linearity assumption of the regression models. Specifically, we estimated the gender and disciplinary differences in partially linear models, while nonparametrically controlling for the effects of potential confounders \cite{chernozhukov2018}. Estimates were both qualitatively and quantitatively similar to the main results from the linear model, providing evidence that observed differences are not driven by a specific modelling approach (Supplementary Materials, Table \ref{tab:s05}).

\section*{Discussion}

This analysis of nearly 40,000 peer review reports submitted to the SNSF from 2016 to 2023 reveals systematic differences in how evaluations are written across disciplines and genders. Reviews in SSH were longer and more critical in tone, with less focus on the track record, whereas those written for MINT proposals were more concise and predominantly positive, with a higher focus on the track record, as compared to those in LS. Female reviewers wrote longer reviews that more closely aligned with the evaluation criteria and expressed more positive sentiments than male reviewers. Female applicants received reviews with slightly more methodological content and a slightly more positive tone. Disciplinary differences shaped both content and sentiment of the reports. MINT reviews emphasized applicants’ track record and feasibility, while in the LS and SSH, reviewers focused more on the methods and the relevance and originality of the proposal. These findings highlight how gender and disciplinary conventions drive the content of grant peer review reports.

Few studies have analyzed peer review reports from funding agencies, and most focused on numerical scores. A landmark 1997 study by the Swedish Medical Research Council found that female applicants for postdoctoral fellowships had to demonstrate greater productivity than males to receive equivalent scores \cite{wenneras1997}. More recently, an analysis of over 38,000 SNSF summary scores showed significant disciplinary variation: fields such as mathematics, physics, geology, chemistry, history, and linguistics received higher scores than medicine, while psychology and sociology received lower or comparable scores \cite{severin2020}. These disciplinary patterns align with our findings from the sentiment analysis. Regarding gender, male reviewers gave higher scores to male applicants, whereas no similar bias was observed among female reviewers \cite{severin2020}. Yet, when moving to the content of reports, we observed that female reviewers wrote more positive and fewer negative comments than their male counterparts. Luo et al. \cite{luo2021}, analyzing reviews from Irish and Swiss funders, found that structured formats and sentiment in specific sections (e.g., feasibility, scientific impact) increased correlation strength. Crucially, in low-success-rate programs, negative sentiment strongly predicted rejection, while in contexts with higher funding rates, sentiment was a weaker indicator \cite{luo2021}. Our study’s 30–40\% success rate reflects a relatively high-funding environment.

Several studies have examined the peer review of journal articles, which—unlike grant peer review—has become more open in recent years. Many journals now publish reviewers’ comments and authors’ responses alongside articles, and platforms have emerged that enable scholars to track their peer review activities and receive recognition for their contributions \cite{godlee2002,ross-hellauer2017}. An analysis of PEERE, a database of almost half a million journal peer review reports \cite{squazzoni2020}, showed that reviewers in the social sciences and economics emphasized methodological rigor more than those in other fields \cite{buljan2024}. Interestingly, our study suggests that reviewers for SSH proposals focused less on the suitability of methods than reviewers for proposals in the LS and MINT. In the PEERE study, the length of the report varied according to the recommendation, with the longest reports supporting a major revision and the shortest recommending acceptance \cite{buljan2024}. Similarly, we found that reports for proposals graded at the bottom or towards the top of the scale received shorter reviews than proposals with intermediate grades. A study of peer review in neuroscience used ChatGPT to examine sentiment and politeness in peer review reports \cite{verharen2023}. Female first authors received less polite reviews than male authors, while female last authors received more favorable ones. The latter aligns with our findings, as last authors would more likely be senior authors and thus more likely eligible to apply for SNSF project funding, and female applicants received more positive reviews.

Strengths of our study include the large and rich dataset from a national funding agency, covering the full range of disciplines and reviewers from across the globe, which allowed for robust comparisons across fields and reviewer and applicant demographics. To our knowledge, this study is one of the first large-scale textual analyses to map the textual content of grant peer review reports. The results not only shed light on the typical length and focus of grant peer reviews, but also uncover important differences across disciplines and the gender of applicants and reviewers. Second, it combines detailed human annotation of review text with fine-tuned transformer-based machine learning models offering high classification accuracy and enabling systematic analysis of review content and sentiment at scale \cite{okasa2024}. In the present analysis, macro F1 scores ranged from 0.79 for \textit{Suitability} to 0.91 for \textit{Track Record}, demonstrating good reliability and balance of false positives and false negatives across most categories \cite{okasa2024}. Third, we developed a transparent, reproducible methodology with publicly shared models, increasing the work’s utility for future research and policy evaluation \cite{okasa2024}. All analyses are validated through human oversight, and the entire classification is conducted locally using openly available methods. This avoids reliance on proprietary models and sidesteps common issues associated with commercial services, such as data privacy concerns, limited transparency and reproducibility, and restricted flexibility \cite{palmer2023}. Fourth, the integration of quantitative (regression) and computational (text classification and keyness analysis) methods allowed nuanced insights into the tone and content of peer review reports. Although the data were drawn from the SNSF, the peer review reports were authored by an international community of scholars. This broad reviewer base increases the likelihood that the findings are generalizable to grant peer review systems in other countries with similar procedures and evaluation structures.

The study also has several limitations. We could not examine whether and how the gender and disciplinary differences observed affect panel discussions and funding rates. Further, our study focused on English-language review reports and is not applicable to the approximately 10\% of the data written in German, French or Italian. The analysis is observational and correlational, limiting the ability to draw causal conclusions about the relationships observed. Although the models adjust for many covariates, unmeasured confounding variables—such as the prestige of the applicant’s institution or prior reviewer-applicant relationships—could still affect the results. Gender is treated as a binary variable, potentially oversimplifying the diversity of reviewer and applicant identities, and only the gender of the corresponding applicant is considered, even though in reality, proposals are often submitted from collaborative teams with a wide variety of gender representations. Finally, while the use of sentiment analysis adds valuable insight into tone, even well-performing machine learning models may struggle to capture subtle or discipline-specific expressions of critique or praise, potentially affecting interpretation \cite{liu2020}.

In conclusion, our study highlights how disciplinary differences and reviewer gender affect the tone and content of grant peer review reports, with potentially important implications for fairness and consistency in funding decisions. The finding that reviewers in different fields emphasize distinct evaluation criteria—such as track record in MINT or methodological rigor in LS—suggests that disciplinary conventions play a role in how scientific merit is operationalized. Similarly, the observation that female reviewers write more positive and detailed evaluations points to gendered communication styles that may affect how funding applications are perceived. These patterns raise concerns about the comparability of peer reviews across disciplines and reviewer demographics, particularly in multidisciplinary panels. The results underscore the need to account for structural variation in review practices when designing peer review systems, setting scoring guidelines, or training reviewers \cite{hren2022,stupacher2025}. Interventions such as reviewer calibration, structured review formats, or diversity policies could help reduce unintended disparities and promote more equitable and transparent evaluation processes. Future research should examine how disciplinary and interdisciplinary panels interpret and integrate review reports with differing emphases or sentiment, and how funding decisions are reached \cite{olbrecht2010}.

\section*{Materials and Methods}

\subsection*{Corpus creation and human annotation}

We restricted our analysis to English-language reviews, using Google’s Compact Language Detector 2, a neural network that detects the language of texts \cite{ooms2025}. The neural network ran locally, ensuring compliance with SNSF’s regulations and Swiss data protection laws. While reviewers in some SSH disciplines can submit their reports in English, German, French, or Italian, most reviews are in English. The full sample of peer review reports submitted between October 2016 and April 2023 contained 44,012 reports written in English, German, French, or Italian, pertaining to 11,977 proposals. We excluded peer review reports written in languages other than English and proposals with an invalid date of birth or missing gender of the applicant. The final analysis dataset included 39,280 peer review reports (89.2\% of total), corresponding to 11,385  proposals, with about 3.5 reviews per proposal. These reports comprised 1,304,621 sentences and 30,477,479 words.

The study combined human annotations of randomly sampled review sentences with fine-tuned transformer-based machine learning models. After preliminary annotation exercises and various inter-coder reliability tests, we finalized an annotation codebook \cite{okasa2024}. Each sentence was randomly assigned to three of the four annotators, resulting in a dataset of 3,000 annotated sentences. Sentences could be assigned to zero, one, or multiple content categories. Final labels used for model training were determined by majority rule. Full agreement, where all three annotators assigned the same label, ranged from 74.1\% (Positive) to 90.7\% (Feasibility). The mean of full agreement across the six categories was 84.5\%. The annotation process is described in detail in Okasa et al. \cite{okasa2024}.

\subsection*{Text classification using transformer models}

We fine-tuned six separate transformer machine learning models, one for each content category. Fine-tuning the pre-trained SPECTER2 model for each content category resulted in good out-of-sample performance. SPECTER2 is a transformer model pre-trained on scientific texts and is suitable for various NLP tasks involving academic texts \cite{singh2023}. The models were fine-tuned using a training set of 2,500 sentences drawn randomly from SNSF grant peer review reports. Model performance was evaluated using a held-out set of 500 sentences and five-fold cross-validation. Fine-tuning was conducted locally, without internet access, to eliminate any risk of data leakage or network interference. All fine-tuned transformer models are openly available from \textit{HuggingFace} at \url{https://huggingface.co/snsf-data}.

The macro F1 scores, which account for class imbalances and are frequently used to evaluate out-of-sample performance \cite{rainio2024}, range from 0.79 (\textit{Suitability of Methods}) to 0.91 (\textit{Track Record}). These scores are in line with, or higher than, those reported in previous attempts to identify content in peer reviews \cite{severin2023}. Table \ref{tab:s06} in the Supplementary Materials reports the performance metrics for the six machine learning models. In addition to fine-tuning transformer models, we also used a locally run Large Language Model, \textsf{Llama 3-8B-Instruct} \cite{ai@meta2024}, however, this approach performed poorly. We therefore opted for the fully reproducible fine-tuned BERT-type model. Finally, we compared separate binary classifiers with a multi-label and multi-task classification. Separate fine-tuned models resulted in better classification performance. Okasa et al. \cite{okasa2024} provide more details on the classification approach and model performance.

\subsection*{Data protection, data sharing and responsible use}

Language detection, fine-tuning, and sentence-level classifications were conducted locally. The publicly shared data do not contain any identifying information about the authors of the review reports or applicants, and neither the code scripts nor the transformer models allow for the extraction of the raw texts of review reports. All authors and human annotators signed data-sharing agreements, and a Data Management Plan was prepared to ensure compliance \cite{strinzel2024}.

While experiments showed that 2,500 annotated sentences were sufficient to produce reliable classification results for a test set of 500 sentences, the sample is still relatively small. As a result, the classification performance should be carefully evaluated before use, and our models must not be deployed for automatic decision-making without human oversight. Additionally, because the training data is limited to a specific source, the models might not perform well for other types of text or languages other than English.

\subsection*{Statistical analysis}

After applying the six fine-tuned classifiers to each sentence to detect the presence or absence of the four evaluation criteria as well as positive and negative sentiment, we aggregated the data to the review level to calculate the prevalence of each content category. Because the categories are not mutually exclusive, a single sentence may be assigned to multiple categories, which can result in combined prevalences exceeding 100\%. This approach, consistent with prior work \cite{severin2023}, enables comparisons of the relative emphasis placed on different aspects within peer review reports.

Given the clustered nature of the data, where each proposal receives multiple reviews, we employ a mixed-effects regression model to examine the relationship between gender, discipline, and review content \cite{gelman2007}. In this model, the effects of reviewer gender, applicant gender, discipline, and control variables are treated as fixed effects, while random intercepts are introduced at the proposal level to account for within-proposal correlation. Formally, we specify the following linear model:

\vspace{-1cm}
\begin{gather*}
Y_{ij} = \alpha + \gamma G_{ij}^R + \delta G_{j}^A + \theta D_{j} + \beta X_{ij} + U_{j} + \varepsilon_{ij} \\
U_{j} \sim \mathcal{N}(0, \sigma_u^2) \\
\varepsilon_{ij} \sim \mathcal{N}(0, \sigma^2)
\end{gather*}

where $Y_{ij}$ represents the text-derived outcome for review $i$ of proposal $j$. $\alpha$ is the overall intercept, $\gamma$ is the coefficient for reviewer $i$’s gender $G_{ij}^R$ evaluating proposal $j$, $\delta$ is the coefficient for applicant $j$’s gender $G_j^A$, $\theta$ is the coefficient for proposal $j$’s discipline $D_j$, and $\beta$ is a vector of fixed coefficients related to the vector of control variables $X_{ij}$, while $U_j$ is the random intercept for proposal $j$ with variance $\sigma_u^2$. The unobserved error term is denoted by $\varepsilon_{ij}$ with variance $\sigma^2$. Our primary interest lies in the estimation of the coefficients $\gamma$, $\delta$, and $\theta$, which measure the gender and disciplinary differences. We use one-hot encoding for the research domain variable $D_j$, leaving out the LS research domain as the reference category.

The models were adjusted for a comprehensive set of potential confounding variables. These include \textit{funding call-specific factors} (e.g., call deadline, with two calls per calendar year) and \textit{applicant-level characteristics}, such as career stage, prior experience with SNSF (whether the applicant had previously received or submitted grants), requested project duration, and type of professorship. We also account for \textit{proposal-level attributes}, including the presence of multiple applicants, resubmission status, collaboration under a lead agency agreement, and classification as use-inspired research. At the \textit{review level}, the models control for whether the reviewer reported the proposal to be within their area of expertise, the length of the review, the number of reminders sent, and the reviewer’s geographic region of affiliation. Finally, \textit{demographic variables}, including applicants’ age were included. All variables used in the regression analysis are detailed in Table \ref{tab_s01} of the Supplementary Materials.

\subsection*{Robustness tests}

We assessed the robustness of our results along three dimensions: \textit{model specification}, \textit{control variables} and \textit{model assumptions}. First, to check the sensitivity of our results to model specification, we additionally test for an interaction between reviewer and applicant gender and estimate the following model:

\vspace{-0.5cm}
\[
\begin{aligned}
Y_{ij} &= \alpha + \gamma G_{ij}^R + \delta G_{j}^A + \zeta (G_{ij}^R \times G_{j}^A) + \theta D_{j} + \beta X_{ij} + U_{j} + \varepsilon_{ij}
\end{aligned}
\]

where the coefficient $\zeta$ measures the gender interaction effect. The results of these analyses are detailed in Supplementary Materials, Table \ref{tab:s03}.

Second, we check the robustness by enlarging the set of control variables by adjusting for the overall grade assigned by the reviewer to a proposal. Until 2021, reviewers graded the proposal on a six-point scale, and a nine-point scale thereafter. We rescale the reviews on a nine-point scale to a scale ranging from 1–6 and use this standardized scale for the robustness test. We estimate the following equation:

\vspace{-0.5cm}
\[
\begin{aligned}
Y_{ij} &= \alpha + \gamma G_{ij}^R + \delta G_{j}^A + \theta D_{j} + \vartheta Z_{ij} + \beta X_{ij} + U_{j} + \varepsilon_{ij}
\end{aligned}
\]

where $\vartheta$ is the coefficient for the grade $Z_{ij}$ given by reviewer $i$ to proposal $j$. The results of the grade analyses are provided in Supplementary Materials, Table \ref{tab:s04}.

Third, we assess the robustness of our results with respect to the linearity assumption of the mixed-effects model. While we keep the gender and disciplinary differences in a linear form, we abandon the linearity assumption for the control variables, which we model fully nonparametrically and learn the functional form of the controls in a data-driven way. In particular, we specify the following partially linear model \cite{robinson1988}:

\vspace{-0.5cm}
\[
\begin{aligned}
Y_{ij} &= \alpha + \gamma G_{ij}^R + \delta G_{j}^A + \theta D_{j} + f(X_{ij}) + \varepsilon_{ij}
\end{aligned}
\]

where $f(\cdot)$ is an unknown function of the control variables.

To estimate the parameters of the above partially linear models, we follow the recent advances in Double Machine Learning \cite{chernozhukov2018}. First, we partial out the coefficients of the policy variables, i.e., gender and discipline, by estimating two nuisance regression models: 1) regressing the outcome on the set of controls and 2) regressing the policy variables on the set of controls. Note that, in practice, we partial out each of the policy variables at a time, see \cite{bach2024}. We estimate these nuisance regressions fully nonparametrically by Random Forests \cite{breiman2001} to allow for any arbitrary functional form. Second, we run a structural residual-on-residual regression via least squares to estimate the coefficients of the policy variables. Such estimated coefficients share the same statistical properties and interpretation as in the linear model. This is achieved by removing the regularization bias via orthogonalization through partialling out, and by removing the overfitting bias via cross-fitting, i.e., estimating the nuisance regressions and the structural regression on separate sample splits \cite{chernozhukov2018}. Within the Double Machine Learning procedure, we also apply clustering on the proposal level \cite{chiang2022}. In addition, due to partialling out each of the policy variables at a time, we apply the Bonferroni correction for inference. We present the results of the partially linear models in Table \ref{tab:s05} in the Supplementary Materials. The substantive conclusions based on the partially linear models mirror the mixed-effects regression results.

\pagebreak

\subsection*{Acknowledgments}

We would like to thank the members of the SNSF Research Council, the SNSF EDI Department and the Scientific Officers from the SNSF for their insightful feedback and constructive comments throughout the development of this work.

\subsection*{Funding} This study was supported by intramural funding at the SNSF. M.E. was supported by special project funding (SNSF grant 32FP30-189498).

\subsection*{Author contributions} 

\noindent Conceptualization: S.M., G.O., M.S., A.J., K.M., and M.E. 

\noindent Funding acquisition: S.M., K.M. and M.E. 

\noindent Formal analysis: S.M. and G.O. 

\noindent Data curation: S.M. and G.O.

\noindent Methodology: G.O., S.M.,  A.J., and M.E. 

\noindent Investigation: S.M., G.O., M.S., A.J., K.M., and M.E. 

\noindent Visualization: S.M. 

\noindent Supervision: S.M., A.J., K.M., and M.E.

\noindent Project administration: M.S., K.M., A.J.

\noindent Resources: S.M., A.J., and M.E. 

\noindent Validation: S.M. and M.E. 

\noindent Writing—original draft: S.M., G.O., and M.E.  

\noindent Writing—review and editing: S.M., G.O., M.S., A.J., K.M., and M.E.

\subsection*{Competing interests} G.O., M.S., A.J., K.M., and M.E. were employed by the SNSF. S.M. has declared that no competing interests exist.

\subsection*{Data and materials availability}  All data required to evaluate the paper’s conclusions are included in the main text and/or Supplementary Materials. Due to data protection laws, in particular the Federal Act on Data Protection (FADP), the texts of grant peer reviews and the review-level dataset cannot be made publicly available \cite{strinzel2024}. 

All fine-tuned transformer models are available at \url{https://huggingface.co/snsf-data}. The code used to conduct the analysis is available in a GitHub repository: \url{https://github.com/snsf-data/ml-gender-disciplines-peer-review}.

\onehalfspacing
\pagebreak
\printbibliography
\doublespacing

\newpage

\appendix

\pagenumbering{arabic}
  \setcounter{figure}{0}
  \renewcommand{\thefigure}{S\arabic{figure}}
  \setcounter{table}{0}
  \renewcommand{\thetable}{S\arabic{table}}
  \setcounter{footnote}{0}

  \pagenumbering{arabic}
  \renewcommand*{\thepage}{A\arabic{page}}
  

\begin{center}

\textbf{\LARGE Supplementary Materials}

\bigskip

{\Large Gender and Discipline Shape Length, Content and Tone  \\  of Grant Peer Review Reports}

\end{center}

\begin{table}[h]
\centering \small
\begin{tabular}{lp{8cm}}
\toprule
\textbf{Variable name} & \textbf{Description} \\
\midrule
gender\_reviewer & gender of reviewer (male/female) \\
\addlinespace
gender\_applicant & gender of (corresponding) applicant. Note: gender (male/female) missing for three applications \\
\addlinespace
ResearchArea & research domain (LS/MINT/SSH) \\
\addlinespace
CallEndDate & submission deadline of call \\
\addlinespace
DurationRequestedMonth & requested duration of the project as submitted by applicant (in months) \\
\addlinespace
IsHasPreviousProjectApproved & indicates if the applicant has previously received a grant from the SNSF \\
\addlinespace
IsHasProjectPartners & indicates if the proposal has been submitted with project partners \\
\addlinespace
IsHasPreviousProjectRequested & indicates if the applicant has previously applied for a grant from the SNSF \\
\addlinespace
IsMultiApplicant & indicates whether application has several applicants. Note: variables on the applicant always rely on characteristics of the corresponding applicant \\
\addlinespace
IsLeadAgencySNF & indicates if the submission is a collaboration with a foreign funder, handled by the SNSF \\
\addlinespace
IsResubmission & indicates if proposal is a resubmission of a previously rejected grant proposal \\
\addlinespace
IsUseInspired & indicates if the project is use-inspired basic research \\
\addlinespace
MatchToApplicationTopic & question whether application topic is within the reviewer's self-reported area of specialization or within reviewer's wider discipline \\
\addlinespace
ntoken\_review & length of review reports (in words) \\
\addlinespace
ReminderCount & number of reminders sent to reviewer \\
\addlinespace
ResearchInstitutionAtProjectStartDateType & research institution type of the applicant \\
\addlinespace
ResponsibleApplicantAgeAtSubmission & age of applicant at time of submission (invalid ages for 59 applications due to data entry errors by the applicants; removed from sample before analysis) \\
\addlinespace
ResponsibleApplicantProfessorshipType & professorship type of the applicant \\
\addlinespace
ReviewerCountry & country of reviewers' institution \\
\addlinespace
ReviewerDegree & highest achieved academic degree of the reviewer \\
\addlinespace
reviewer\_region & region of reviewer (recoded from ReviewerCountry) \\
\bottomrule
\end{tabular}
\caption{Overview of variables used in analysis}
\label{tab_s01}
\end{table}

\clearpage

\begin{table}[h]
\centering \small
\begin{tabular}{p{8cm}p{7cm}}
\toprule
\textbf{Full description of SNSF discipline group} & \textbf{Short label used in manuscript} \\
\midrule
Art Studies; Musicology; Theatre and Film Studies; Architecture & Art and Design \\
\addlinespace
Astronomy; Astrophysics and Space Sciences & Astronomy \\
\addlinespace
Basic Biological Research & Basic Biological Research \\
\addlinespace
Basic Medical Sciences & Basic Medical Sciences \\
\addlinespace
Chemistry & Chemistry \\
\addlinespace
Clinical Medicine & Clinical Medicine \\
\addlinespace
Earth Sciences & Earth Sciences \\
\addlinespace
Economics; Law & Economics and Law \\
\addlinespace
Engineering Sciences & Engineering and Informatics \\
\addlinespace
Environmental Sciences & \\
\addlinespace
Ethnology; Social and Human Geography & Ethnology and Social Geography \\
\addlinespace
Experimental Medicine & Experimental Medicine \\
\addlinespace
General Biology & General Biology \\
\addlinespace
Linguistics and Literature; Philosophy & Linguistics, Literature, and Philosophy \\
\addlinespace
Mathematics & Mathematics \\
\addlinespace
Physics & Physics \\
\addlinespace
Preventive Medicine (Epidemiology/Early Diagnosis/Prevention) & Epidemiology and Preventive Medicine \\
\addlinespace
Psychology; Educational Studies & Psychology and Educational Studies \\
\addlinespace
Social Medicine & Social Medicine \\
\addlinespace
Sociology; Social Work; Political Sciences; Media and Communication Studies; Health & Social and Political Sciences \\
\addlinespace
Theology and Religious Studies; History; Classical Studies; Archaeology & Historical and Religious Studies \\
\bottomrule
\end{tabular}
\caption{Comparison of 21 SNSF disciplines and the more concise descriptions used in the main text and graphs. See classification of disciplines here: \url{https://www.snf.ch/SiteCollectionDocuments/allg_disziplinenliste.pdf}}
\label{tab_s02}
\end{table}

\clearpage

\begin{figure}[ht]
\centering
\includegraphics[width=1\textwidth]{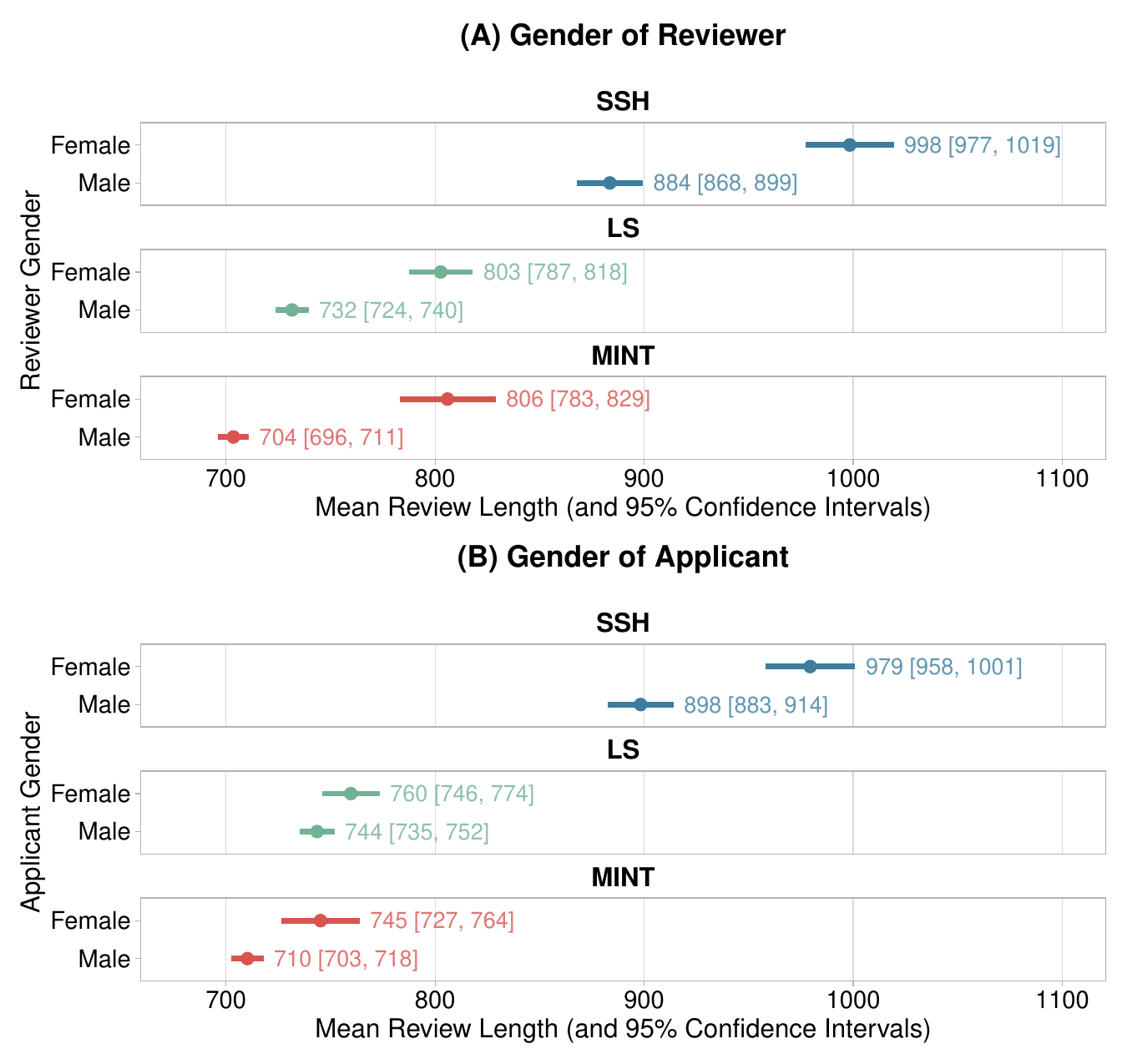}
\caption{\textit{Average review length for reports submitted by male and female reviewers (A) and male and female applicants (B)}. Dots show mean length (in words), horizontal bars depict 95\% confidence intervals. SSH: Social Sciences and Humanities; LS: Life Sciences; MINT: Mathematics, Informatics, Natural Sciences, Technology.} \label{fig_s01}
\end{figure}

\begin{figure}[ht]
\centering
\includegraphics[width=1\textwidth]{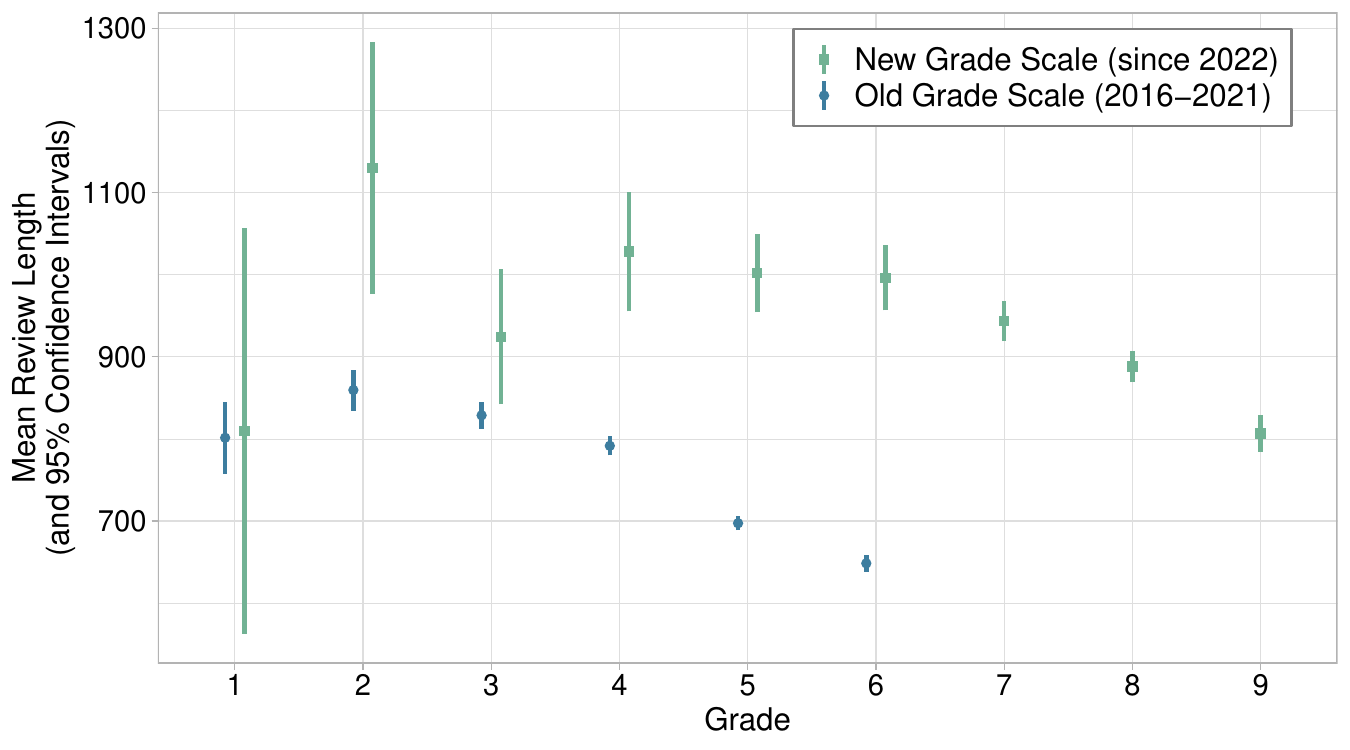}
\caption{\textit{Relationship between grade and review length}. Higher values imply better grades assigned by the reviewer to a proposal. The grades between 2016 and 2021 (blue color) rely on a six-point scale; the more recent grades, introduced in 2022 (green color), rely on a nine-point scale.} \label{fig_s02}
\end{figure}

\begin{figure}[ht]
\centering
\includegraphics[width=1\textwidth]{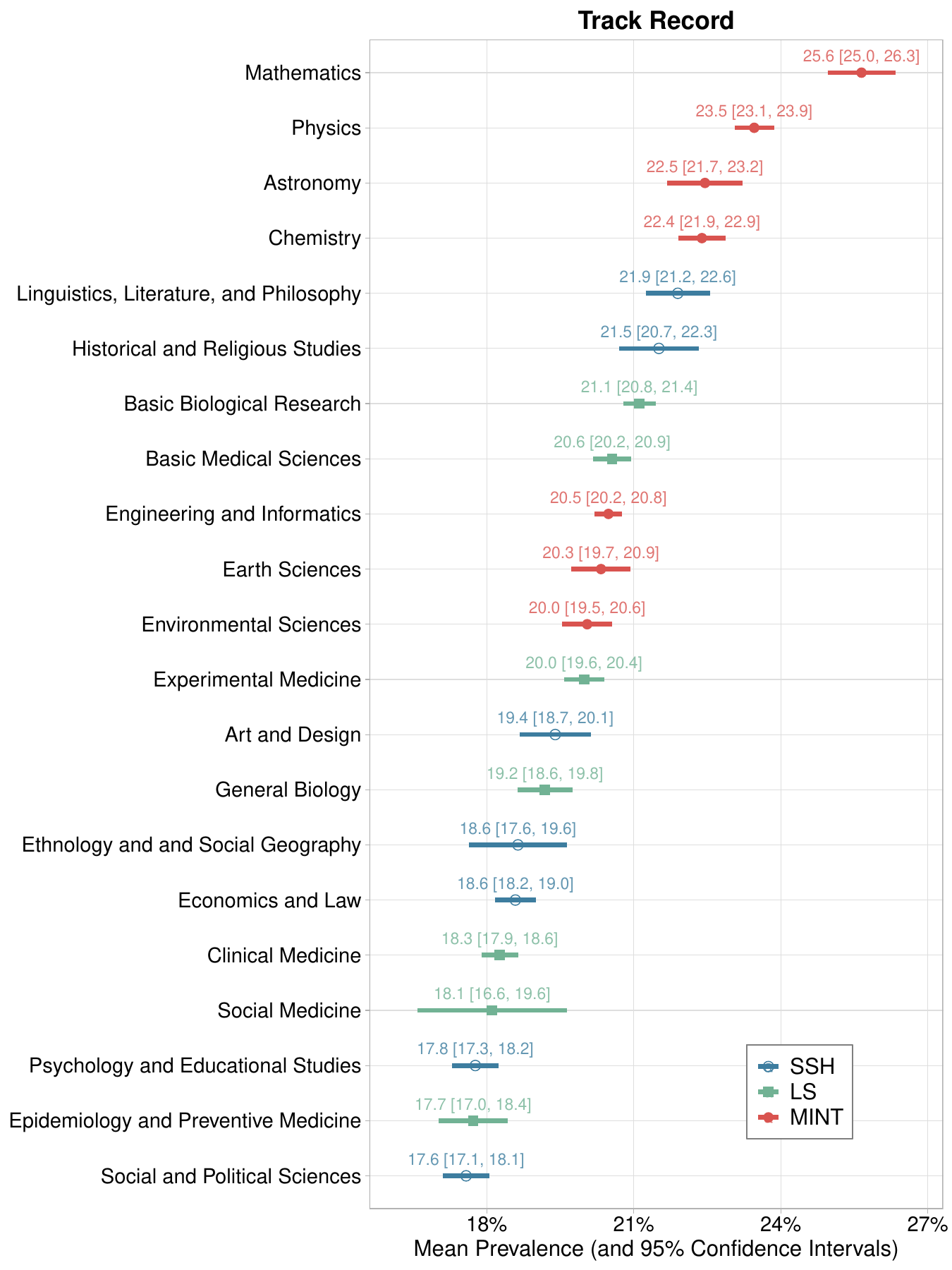}
\caption{\textit{Mean prevalence of Track Record (\% of review sentences) by discipline}. Horizontal bars depict 95\% confidence intervals. Mathematics focuses most on the Applicant's Track Record (25.6\%), followed by Physics (23.5\%), and Astronomy (22.5\%). On the other hand, Social and Political Sciences place the least emphasis on the track record (17.6\%).} \label{fig_s03}
\end{figure}

\begin{figure}[ht]
\centering
\includegraphics[width=1\textwidth]{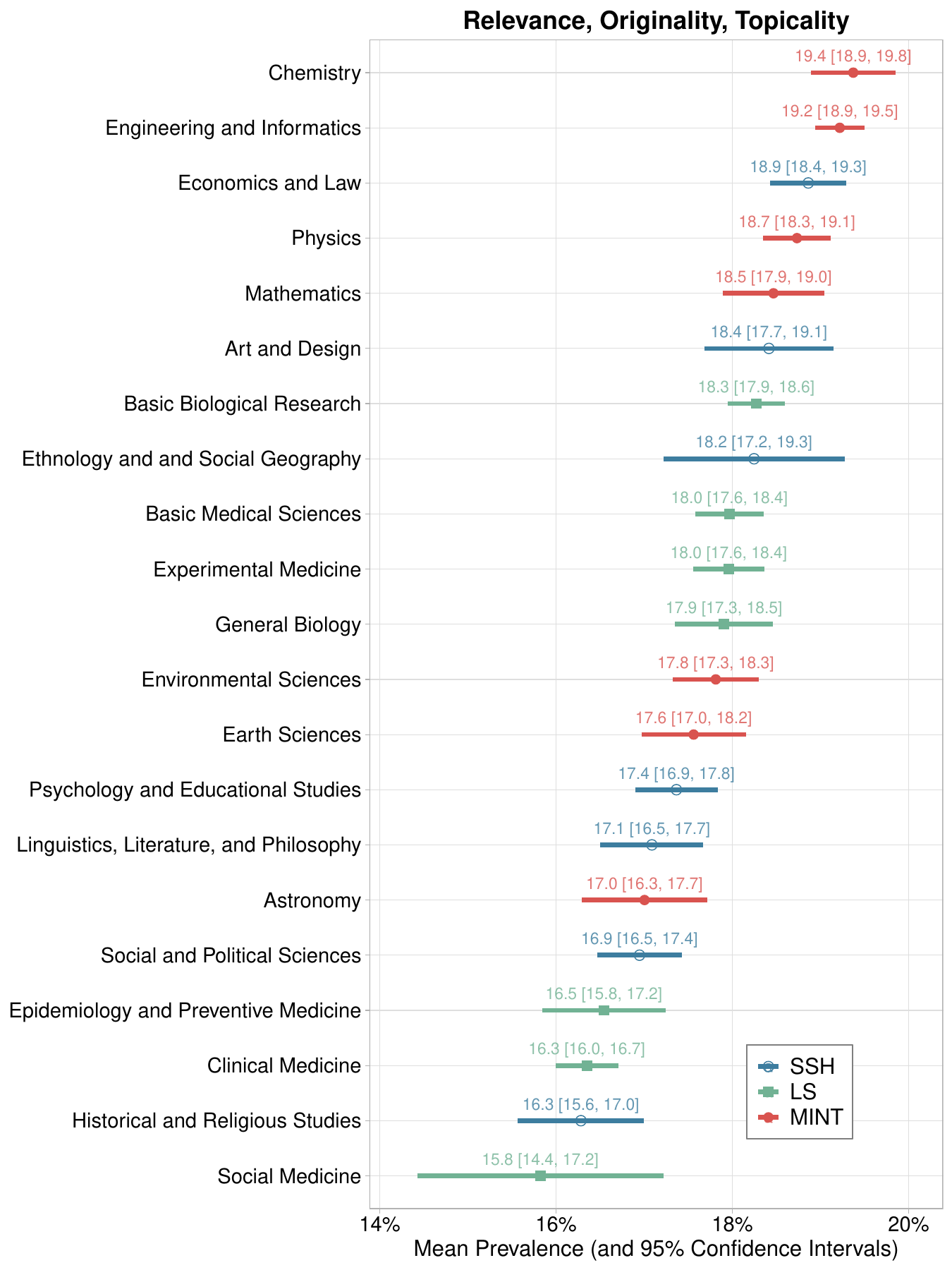}
\caption{\textit{Mean prevalence of Relevance, Originality, Topicality (\% of review sentences) by discipline}. Horizontal bars depict 95\% confidence intervals. Mean differences in the focus on Relevance, Originality, and Topicality tend to be smaller than in other research areas (ranging from 15.8\% to 19.4\%), and we do not observe clusters across the three research domains.} \label{fig_s04}
\end{figure}

\begin{figure}[ht]
\centering
\includegraphics[width=1\textwidth]{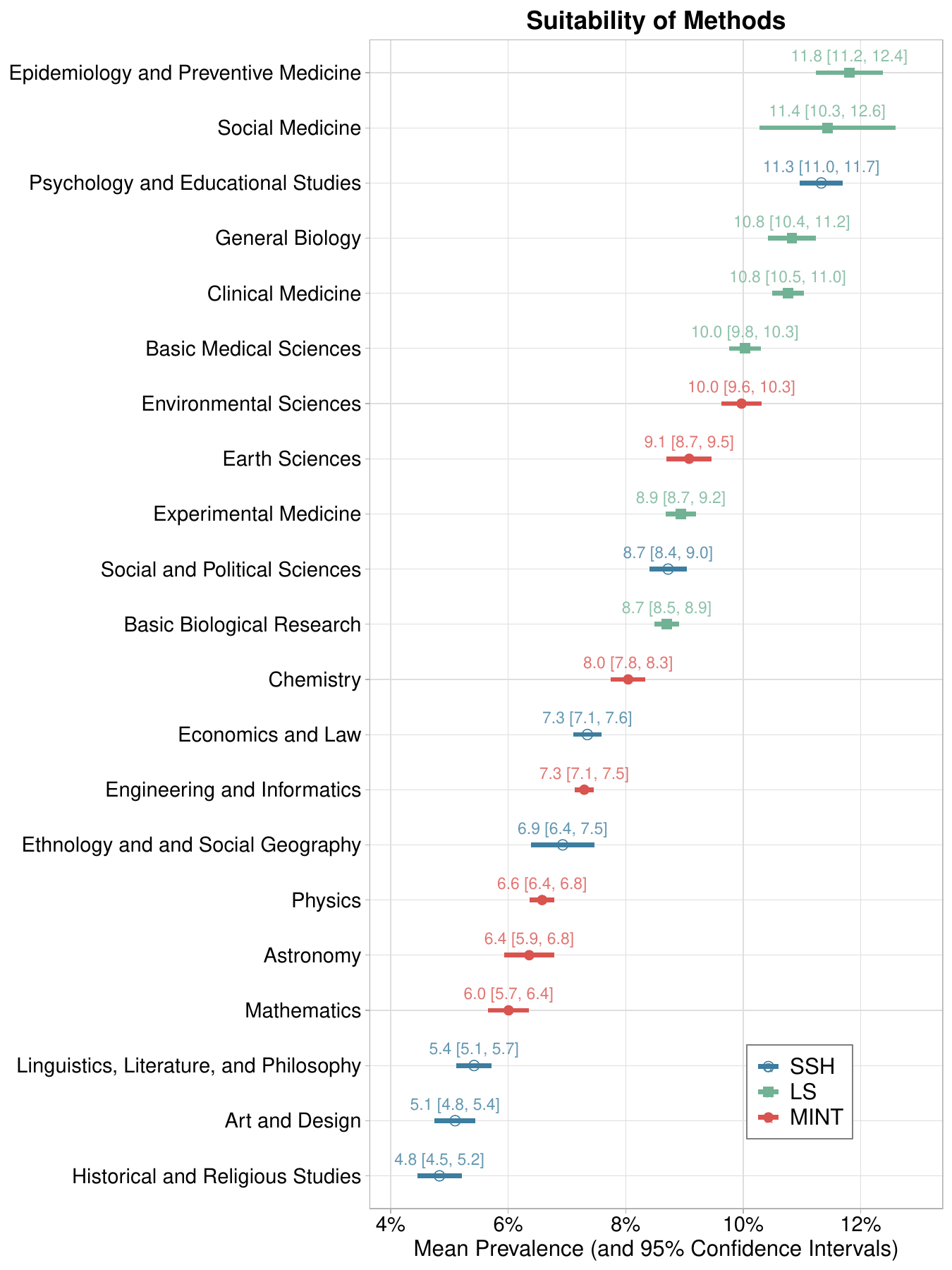}
\caption{\textit{Mean prevalence of Suitability of Methods (\% of review sentences) by discipline}. Horizontal bars depict 95\% confidence intervals. Reviews in the Life Sciences devote considerably more emphasis on the Suitability of Methods (11.8\% in Preventive Medicine, 11.4\% in Social Medicine) than SSH disciplines, in particular Historical and Religious Studies (4.8\%) and Art and Design (5.1\%).} \label{fig_s05}
\end{figure}

\begin{figure}[ht]
\centering
\includegraphics[width=1\textwidth]{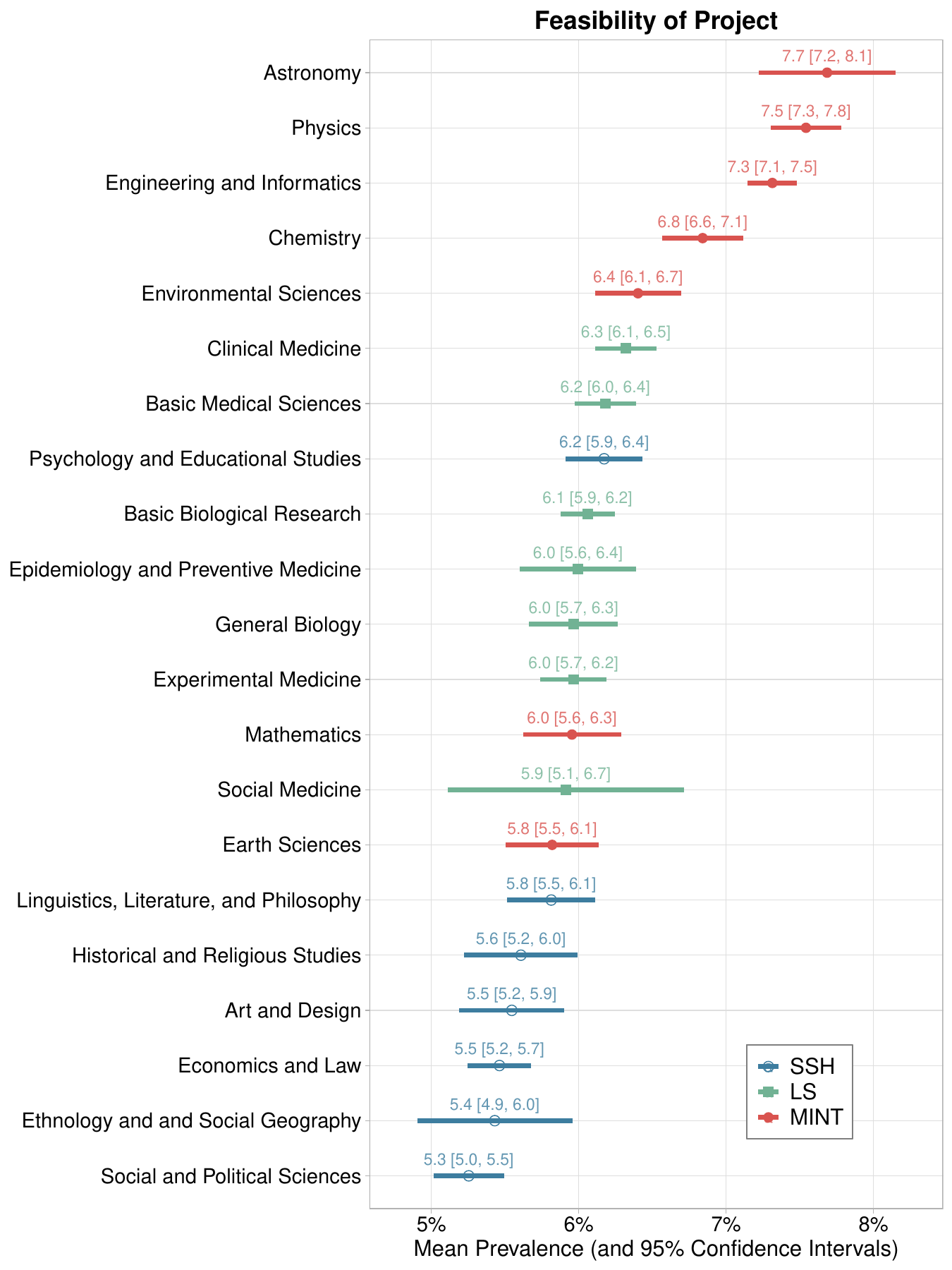}
\caption{\textit{Mean prevalence of Feasibility of Project (\% of review sentences) by discipline}. Horizontal bars depict 95\% confidence intervals. The feasibility of the project is discussed somewhat more extensively in MINT and LS than in SSH, even though these differences are relatively small, ranging from 5.3\% in Social and Political Sciences to 7.7\% in Astronomy.} \label{fig_s06}
\end{figure}

\begin{landscape}
\begin{table}
\centering
\begin{talltblr}[         
caption={\textit{Predicting the prevalence of review characteristics by considering interaction effects between the gender of the reviewer and applicant}. Content categories are listed in the first row. Table reports results based on mixed-effects regression models. Coefficients show percentage point differences in prevalence relative to reference category. Models include control variables listed in Table \ref{tab_s01} and random intercepts for proposal IDs. 95\% confidence intervals in parentheses.\label{tab:s03}},
]                     
{                     
colspec={Q[]Q[]Q[]Q[]Q[]Q[]Q[]},
cell{1}{2}={c=4,}{halign=c,},
cell{1}{6}={c=2,}{halign=c,},
column{1}={halign=l,},
column{2}={halign=c,},
column{3}={halign=c,},
column{4}={halign=c,},
column{5}={halign=c,},
column{6}={halign=c,},
column{7}={halign=c,},
row{2}={font=\fontsize{1em}{1.3em}\selectfont,},
row{1}={font=\fontsize{1em}{1.3em}\selectfont,},
row{3}={font=\fontsize{1em}{1.3em}\selectfont,},
row{4}={font=\fontsize{1em}{1.3em}\selectfont,},
row{5}={font=\fontsize{1em}{1.3em}\selectfont,},
row{6}={font=\fontsize{1em}{1.3em}\selectfont,},
row{7}={font=\fontsize{1em}{1.3em}\selectfont,},
row{8}={font=\fontsize{1em}{1.3em}\selectfont,},
row{9}={font=\fontsize{1em}{1.3em}\selectfont,},
row{10}={font=\fontsize{1em}{1.3em}\selectfont,},
row{11}={font=\fontsize{1em}{1.3em}\selectfont,},
row{12}={font=\fontsize{1em}{1.3em}\selectfont,},
}                     
\toprule
& Evaluation Criteria &  &  &  & Sentiment &  \\ \cmidrule[lr]{2-5}\cmidrule[lr]{6-7}
& Track Record & Rel., Orig., Topic. & Methods & Feasibility & Positive & Negative \\ \midrule 
Reviewer: Female (ref.: Male)        & 0.87           & 0.70          & 0.50           & 0.50           & 0.28          & -0.63          \\
& [0.56, 1.18]   & [0.41, 1.00]  & [0.30, 0.70]   & [0.33, 0.67]   & [-0.25, 0.82] & [-1.03, -0.24] \\
Applicant: Female (ref.: Male)       & 0.11           & -0.11         & 0.45           & -0.06          & 0.23          & -0.15          \\
& [-0.20, 0.41]  & [-0.40, 0.18] & [0.25, 0.66]   & [-0.23, 0.11]  & [-0.31, 0.77] & [-0.56, 0.25]  \\
Domain: MINT (ref.: LS)              & 1.81           & 0.60          & -2.19          & 0.73           & 4.63          & -1.70          \\
& [1.51, 2.10]   & [0.33, 0.88]  & [-2.39, -1.99] & [0.57, 0.89]   & [4.10, 5.15]  & [-2.09, -1.30] \\
Domain: SSH (ref.: LS)               & -0.78          & 0.80          & -2.09          & -0.45          & 2.68          & 1.62           \\
& [-1.10, -0.46] & [0.50, 1.11]  & [-2.31, -1.87] & [-0.62, -0.27] & [2.11, 3.25]  & [1.19, 2.05]   \\
Reviewer: Female x Applicant: Female & 0.42           & 0.19          & -0.07          & 0.32           & 1.00          & -1.13          \\
& [-0.12, 0.96]  & [-0.32, 0.71] & [-0.42, 0.28]  & [0.02, 0.62]   & [0.06, 1.93]  & [-1.83, -0.44] \\
\bottomrule
\end{talltblr}
\end{table}

\end{landscape}

\begin{figure}[ht]
\centering
\includegraphics[width=1\textwidth]{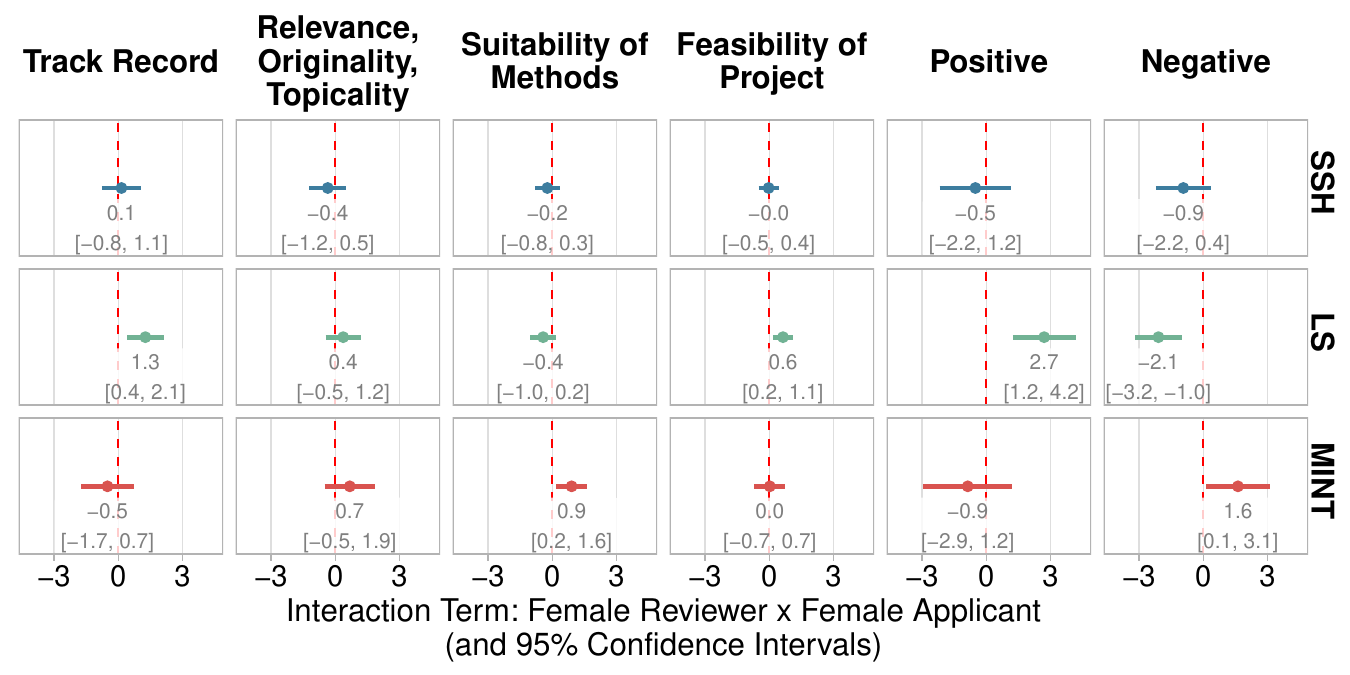}
\caption{\textit{Interaction terms for female reviewers and female applicants}. Each box shows the interaction term for a separate model and subset of the dataset for each research area, resulting in 18 regression models. All regression models include the control variables described in the main paper. Horizontal bars depict 95\% confidence intervals.
} \label{fig_s07}
\end{figure}

\begin{landscape}

\begin{table}
\centering
\begin{talltblr}[         
caption={\textit{Predicting the prevalence of review characteristics by the grade assigned by the reviewer}. Content categories are listed in the first row. Table reports results based on mixed-effects regression models. The variable Grade (rescaled) rescales the new grade scale from a nine- to a six-point scale to facilitate comparability across the full sample. Coefficients show percentage point differences in prevalence relative to reference category. Models include control variables listed in Table \ref{tab_s01} and random intercepts for proposal IDs. 95\% confidence intervals in parentheses. Track Record: Applicant's Track Record; Rel., Orig., Topic: Relevance, Originality, Topicality; Methods: Suitability of Methods; Feasibility: Feasibility of Project.\label{tab:s04}},
]                     
{                     
colspec={Q[]Q[]Q[]Q[]Q[]Q[]Q[]},
cell{1}{2}={c=4,}{halign=c,},
cell{1}{6}={c=2,}{halign=c,},
column{1}={halign=l,},
column{2}={halign=c,},
column{3}={halign=c,},
column{4}={halign=c,},
column{5}={halign=c,},
column{6}={halign=c,},
column{7}={halign=c,},
row{2}={font=\fontsize{1em}{1.3em}\selectfont,},
row{1}={font=\fontsize{1em}{1.3em}\selectfont,},
row{3}={font=\fontsize{1em}{1.3em}\selectfont,},
row{4}={font=\fontsize{1em}{1.3em}\selectfont,},
row{5}={font=\fontsize{1em}{1.3em}\selectfont,},
row{6}={font=\fontsize{1em}{1.3em}\selectfont,},
row{7}={font=\fontsize{1em}{1.3em}\selectfont,},
row{8}={font=\fontsize{1em}{1.3em}\selectfont,},
row{9}={font=\fontsize{1em}{1.3em}\selectfont,},
row{10}={font=\fontsize{1em}{1.3em}\selectfont,},
row{11}={font=\fontsize{1em}{1.3em}\selectfont,},
row{12}={font=\fontsize{1em}{1.3em}\selectfont,},
}                     
\toprule
& Evaluation Criteria &  &  &  & Sentiment &  \\ \cmidrule[lr]{2-5}\cmidrule[lr]{6-7}
& Track Record & Rel., Orig., Topic. & Methods & Feasibility & Positive & Negative \\ \midrule 
Reviewer: Female (ref.: Male)  & 1.06           & 0.83          & 0.48           & 0.62           & 0.82         & -1.11          \\
& [0.81, 1.31]   & [0.59, 1.07]  & [0.31, 0.64]   & [0.48, 0.76]   & [0.46, 1.19] & [-1.37, -0.86] \\
Applicant: Female (ref.: Male) & 0.20           & -0.08         & 0.44           & 0.02           & 0.37         & -0.33          \\
& [-0.05, 0.45]  & [-0.32, 0.16] & [0.26, 0.63]   & [-0.12, 0.17]  & [0.01, 0.73] & [-0.59, -0.08] \\
Domain: MINT (ref.: LS)        & 1.44           & 0.26          & -2.12          & 0.66           & 3.18         & -0.48          \\
& [1.17, 1.72]   & [-0.01, 0.52] & [-2.32, -1.92] & [0.50, 0.81]   & [2.78, 3.57] & [-0.76, -0.20] \\
Domain: SSH (ref.: LS)         & -0.67          & 0.90          & -2.10          & -0.41          & 3.01         & 1.34           \\
& [-0.98, -0.37] & [0.60, 1.19]  & [-2.32, -1.89] & [-0.59, -0.24] & [2.58, 3.45] & [1.03, 1.64]   \\
Grade (rescaled from 1-6)      & 2.49           & 2.33          & -0.45          & 0.54           & 9.60         & -8.00          \\
& [2.40, 2.58]   & [2.24, 2.41]  & [-0.51, -0.39] & [0.48, 0.59]   & [9.46, 9.73] & [-8.10, -7.91] \\
\bottomrule
\end{talltblr}
\end{table}

\pagebreak
\newpage 

\begin{table}
\centering
\begin{talltblr}[         
caption={\textit{Predicting the prevalence of review characteristics in a partially linear model.} Content categories are listed in the first row. Table reports results based on partially linear 
regression models estimated via double machine learning with random forest with 1,000 trees as a base learner and 5-fold cross-fitting. Coefficients show percentage point differences in prevalence relative to reference category. Models include control variables listed in  Table \ref{tab_s01} and clustering on proposal IDs. 95\% confidence intervals with Bonferroni correction in parentheses. Track Record: Applicant's Track Record; Rel., Orig., Topic: Relevance, Originality, Topicality; Methods: Suitability of Methods; Feasibility: Feasibility of Project.\label{tab:s05}},
]                     
{                     
colspec={Q[]Q[]Q[]Q[]Q[]Q[]Q[]},
cell{1}{2}={c=4,}{halign=c,},
cell{1}{6}={c=2,}{halign=c,},
column{1}={halign=l,},
column{2}={halign=c,},
column{3}={halign=c,},
column{4}={halign=c,},
column{5}={halign=c,},
column{6}={halign=c,},
column{7}={halign=c,},
row{2}={font=\fontsize{1em}{1.3em}\selectfont,},
row{1}={font=\fontsize{1em}{1.3em}\selectfont,},
row{3}={font=\fontsize{1em}{1.3em}\selectfont,},
row{4}={font=\fontsize{1em}{1.3em}\selectfont,},
row{5}={font=\fontsize{1em}{1.3em}\selectfont,},
row{6}={font=\fontsize{1em}{1.3em}\selectfont,},
row{7}={font=\fontsize{1em}{1.3em}\selectfont,},
row{8}={font=\fontsize{1em}{1.3em}\selectfont,},
row{9}={font=\fontsize{1em}{1.3em}\selectfont,},
row{10}={font=\fontsize{1em}{1.3em}\selectfont,},
}                     
\toprule
& Evaluation Criteria &  &  &  & Sentiment &  \\ \cmidrule[lr]{2-5}\cmidrule[lr]{6-7}
& Track Record &  Rel., Orig., Topic.  & Methods & Feasibility & Positive & Negative \\ \midrule 
Reviewer: Female (ref.: Male) & 0.98 & 0.86 & 0.66 & 0.62 & 0.83 & -0.86 \\ 
  & [0.64; 1.31] & [0.55; 1.18] & [0.44; 0.88] & [0.44; 0.8] & [0.26; 1.4] & [-1.29; -0.43] \\[1.5ex] 
 Applicant: Female (ref: Male) & 0.24 & -0.04 & 0.46 & 0.01 & 0.49 & -0.35 \\ 
  & [-0.1; 0.58] & [-0.36; 0.28] & [0.23; 0.7] & [-0.17; 0.2] & [-0.09; 1.08] & [-0.79; 0.09] \\[1.5ex] 
  Domain: MINT (ref.: LS) & 1.7 & 0.46 & -2.16 & 0.65 & 4.13 & -1.66 \\ 
 & [1.31; 2.09] & [0.08; 0.83] & [-2.43; -1.88] & [0.43; 0.87] & [3.45; 4.81] & [-2.18; -1.14] \\[1.5ex] 
  Domain: SSH (ref.: LS) & -0.7 & 0.86 & -2.16 & -0.47 & 2.31 & 1.73 \\ 
   & [-1.15; -0.26] & [0.45; 1.28] & [-2.47; -1.84] & [-0.7; -0.24] & [1.52; 3.09] & [1.13; 2.34] \\ 
 \bottomrule
\end{talltblr}
\end{table}

\pagebreak
\newpage 

\begin{table}
\centering 
\begin{talltblr}[         
caption={\textit{Overview of the classification test results based on binary classification.} Table shows Accuracy (Acc.), Balanced Accuracy (Bal. Acc.), the F1 score, Precision (Prec.), Recall (Rec.), and class-specific performance metrics for F1 score, Precision, and Recall for the presence (Lab=1) and absence (Lab=0) of a characteristic. Results are based on six separate fine-tuned transformer models. The metrics can range from 0 to 1 with higher values indicating better model classification. Detailed results and systematic comparisons of various classification approaches are presented in Okasa et al. \cite{okasa2024}.\label{tab:s06}},
]                     
{                     
colspec={Q[]Q[]Q[]Q[]Q[]Q[]Q[]Q[]Q[]Q[]Q[]Q[]},
row{1}={font=\fontsize{1em}{1.3em}\selectfont,},
row{2}={font=\fontsize{1em}{1.3em}\selectfont,},
row{3}={font=\fontsize{1em}{1.3em}\selectfont,},
row{4}={font=\fontsize{1em}{1.3em}\selectfont,},
row{5}={font=\fontsize{1em}{1.3em}\selectfont,},
row{6}={font=\fontsize{1em}{1.3em}\selectfont,},
row{7}={font=\fontsize{1em}{1.3em}\selectfont,},
}                     
\toprule
Category & Acc. & \makecell{Bal. \\ Acc.} & F1 & \makecell{F1 \\ Lab=1} & \makecell{F1 \\ Lab=0} & Prec. & \makecell{Prec. \\ Lab=1} & \makecell{Prec. \\ Lab=0} & Rec. & \makecell{Rec. \\ Lab=1} & \makecell{Rec. \\ Lab=0} \\ \midrule 
Track Record                       & 0.95 & 0.92 & 0.91 & 0.86 & 0.97 & 0.90 & 0.82 & 0.98 & 0.92 & 0.89 & 0.96 \\
Rel., Orig., Topic. & 0.92 & 0.86 & 0.86 & 0.76 & 0.95 & 0.85 & 0.75 & 0.95 & 0.86 & 0.78 & 0.95 \\
Suitability of Methods             & 0.94 & 0.77 & 0.79 & 0.62 & 0.97 & 0.82 & 0.69 & 0.96 & 0.77 & 0.56 & 0.98 \\
Feasibility of Project             & 0.97 & 0.85 & 0.87 & 0.76 & 0.99 & 0.90 & 0.81 & 0.98 & 0.85 & 0.71 & 0.99 \\
Positive                           & 0.90 & 0.90 & 0.89 & 0.87 & 0.92 & 0.89 & 0.84 & 0.94 & 0.90 & 0.90 & 0.90 \\
Negative                           & 0.91 & 0.84 & 0.83 & 0.71 & 0.94 & 0.81 & 0.67 & 0.95 & 0.84 & 0.75 & 0.93 \\
\bottomrule
\end{talltblr}
\end{table}

\end{landscape}

\end{document}